\documentclass[12pt]{article}

\usepackage{scicite}
\usepackage{graphicx}
\usepackage{times}

\newenvironment{sciabstract}{%
\begin{quote} \bf}
{\end{quote}}

\title{
Tunable Isolated Attosecond X-ray Pulses with Gigawatt Peak Power from a Free-Electron Laser} 
\author{
Joseph Duris$^{*1}$, Siqi Li$^{*1,2}$, Taran Driver$^{1,4,5}$, Elio G. Champenois$^{4}$, \\James P. MacArthur$^{1,2}$, Alberto A. Lutman$^{1}$, Zhen Zhang$^{1}$, Philipp Rosenberger$^{1,4,6,7}$,\\Jeff W. Aldrich$^{1}$,  Ryan Coffee$^{1,4}$, Giacomo Coslovich$^{1}$, Franz-Josef Decker$^{1}$,\\ James M. Glownia$^{1,4}$, Gregor Hartmann$^{8}$, Wolfram Helml$^{7,9,10}$,  Andrei Kamalov$^{2,4}$,\\Jonas Knurr$^{4}$, Jacek Krzywinski$^{1}$, Ming-Fu Lin$^{1}$, Megan Nantel$^{1,2}$, Adi Natan$^{4}$,\\ Jordan O'Neal$^{2,4}$, Niranjan Shivaram$^{1}$, Peter Walter$^{1}$, Anna Wang$^{3,4}$,\\ James J. Welch$^{1}$, Thomas J. A. Wolf$^{4}$, Joseph Z. Xu$^{11}$, \\Matthias F. Kling$^{1,4,6,7}$, Philip~H.~Bucksbaum$^{1,2,3,4}$, Alexander Zholents$^{11}$, \\Zhirong Huang$^{1,3}$, James P. Cryan$^{1,4,\dagger}$, and Agostino Marinelli$^{1,\dagger}$
\\    
\normalsize{$^{1}$SLAC National Accelerator Laboratory, Menlo Park, CA, 94025, USA}\\
\normalsize{$^{2}$Physics Department, Stanford University, Stanford, CA, 94305, USA}\\
\normalsize{$^{3}$Applied Physics Department, Stanford University, Stanford, CA, 94305, USA}\\
\normalsize{$^{4}$Stanford PULSE Institute, SLAC National Accelerator Laboratory, Menlo Park, CA, 94025, USA}\\
\normalsize{$^{5}$The Blackett Laboratory, Imperial College, London, SW7 2AZ, UK}\\
\normalsize{$^{6}$ Max Planck Institute of Quantum Optics, D-85748 Garching, Germany}\\
\normalsize{$^{7}$ Physics Department, Ludwig-Maximilians-Universit\"at Munich, 85748 Garching, Germany}\\
\normalsize{$^{8}$Institut f\"ur Physik und CINSaT, Universit\"at Kassel, Heinrich-Plett-Str. 40, 34132 Kassel, Germany 
}\\
\normalsize{$^{9}$ Zentrum f\"ur Synchrotronstrahlung, Technische Universit\"at Dortmund,}\\ \normalsize{Maria-Goeppert-Mayer-Stra\ss e 2, 44227 Dortmund, Germany} \\
\normalsize{$^{10}$ Physik-Department E11, Technische Universit\"at M\"unchen,}\\
\normalsize{James Franck-Stra\ss e 1, 85748 Garching, Germany}\\
\normalsize{$^{11}$Argonne National Laboratory, Lemont, IL, 60439, USA}\\
\normalsize{$^{*}$These authors contributed equally to this work.}\\
\normalsize{$^{\dagger}$To whom correspondence should be addressed;}\\ 
\normalsize{E-mail: marinelli@slac.stanford.edu,   jcryan@slac.stanford.edu.}\\\\\\
}

\date{}

\begin{document} 


\baselineskip24pt


\maketitle


\begin{sciabstract}
The quantum mechanical motion of electrons in molecules and solids occurs on the sub-femtosecond timescale.  
Consequently, the study of ultrafast electronic phenomena requires the generation of laser pulses shorter than 1~fs and of sufficient intensity to interact with their target with high probability. 
Probing these dynamics with atomic-site specificity requires the extension of sub-femtosecond pulses to the soft X-ray spectral region.
Here we report the generation of isolated GW-scale soft X-ray attosecond pulses with an X-ray free-electron laser.
Our source has a pulse energy that is six orders of magnitude larger than any other source of isolated attosecond pulses in the soft X-ray spectral region, with a peak power in the tens of gigawatts. 
This unique combination of high intensity, high photon energy and short pulse duration enables the investigation of electron dynamics with X-ray non-linear spectroscopy and single-particle imaging.
\end{sciabstract}

\section*{Introduction}
The natural time scale of electron motion in molecular systems is determined by the binding energy, $I_p$, typically between 8~and~12~eV. 
Quantum mechanics tells us that this relationship is given by  $\tau = \hbar / I_p$, where $\hbar$ is the reduced Planck constant.
Therefore, the relevant time scale for electron motion in molecular systems is on the order of a few hundred attoseconds (1~as$~=10^{-18}$~sec). 
Light pulses approaching this extreme timescale were first demonstrated in 2001~\cite{hentschel_attosecond_2001}.
These early demonstrations employed a process called high harmonic generation~(HHG), where a strong, infrared laser-field was used to coherently drive electrons in an atomic or molecular gas, leading to high-order harmonic up-conversion of the driving laser field~\cite{li_multiple-harmonic_1989,krause_high-order_1992,corkum_plasma_1993,chini_generation_2014}.   
The extension of time-resolved spectroscopy into the attosecond domain has greatly advanced our understanding of electron dynamics in atoms, molecules, and condensed matter systems~\cite{corkum_attosecond_2007,chang_attosecond_2010,ciappina_attosecond_2017}. 
This attosecond revolution has been almost exclusively driven by HHG based sources~\cite{hentschel_attosecond_2001,sekikawa_nonlinear_2004,sansone_isolated_2006,sola_controlling_2006,goulielmakis_single-cycle_2008,mashiko_double_2008,feng_generation_2009,ferrari_high-energy_2010,takahashi_attosecond_2013,ossiander_attosecond_2017,barillot_towards_2017,bergues_tabletop_2018,jahn_towards_2019}, which have been recently extended to reach the soft X-ray wavelengths (above 280~eV~\cite{AttwoodBook}) and produce the shortest pulses ever recorded~\cite{teichmann_0.5-kev_2016,gaumnitz_streaking_2017,li_53-attosecond_2017,johnson_high-flux_2018}.
Extending attosecond pulse sources into the soft X-ray domain is particularly important because soft X-rays can access core-level electrons whose absorption properties are sensitive probes of transient electronic structure~\cite{wolf_probing_2017,neville_ultrafast_2018,attar_femtosecond_2017}.

In parallel with the development of HHG, the last two decades have seen the rise of X-ray free-electron lasers~(XFELs), such as the Linac Coherent Light Source~(LCLS), as the brightest sources of X-ray radiation~\cite{FLASHnp,LCLS,FERMInp,FERMI2stage,SACLA,POHANG,altarelli2011european}.
The working principle of an XFEL is based on the interaction of a relativistic electron beam with an X-ray electric field in a long periodic array of magnetic dipoles called an undulator~\cite{BPN,ClaudioRMP,HuangKimReview}.
The radiation/electron interaction causes the electron beam to re-organize itself in a sequence of microbunches shorter than the radiation wavelength, which results in the coherent emission of X-ray radiation with a peak power many orders of magnitude larger than the spontaneous level~\cite{ClaudioRMP,HuangKimReview}.
Compared to laser-based HHG sources XFELs have a large extraction efficiency at X-ray wavelengths, typically of order 0.1\% or larger. 
With a typical electron beam peak power in the tens of terawatts range, the resulting X-ray pulses have tens of gigawatts of peak power, several orders of magnitude larger than table-top X-ray sources.  
Furthermore, the photon energy of FEL sources is easily tunable via small configuration changes of the accelerator or the undulator.
The shortest pulse achievable with an XFEL is limited by the available amplification bandwidth, which is of similar magnitude to their extraction efficiency $\leq 0.1\%$~\cite{BPN,PellegriniSpectrum}.
For example, the X-ray bandwidth of LCLS can support pulses shorter than 1~fs for hard X-ray energies~\cite{huang2017generating,marinelli2017experimental}.
However, the shortest possible pulse duration increases to 1-2~fs for photon energies below 1~keV \cite{behrens_few-femtosecond_2014,hartmann_attosecond_2018}, where the relevant core-level absorption edges for light elements are found: carbon~(280~eV), nitrogen~(410~eV), and oxygen~(540~eV).
In this letter, we report the generation and time-resolved measurement of gigawatt-scale isolated attosecond soft X-ray pulses with an XFEL. 
The bandwidth limitation of the XFEL was overcome by compressing the electron beam with a high-power infrared pulse, a technique termed enhanced self-amplified spontaneous emission~(ESASE)~\cite{zholents_method_2005}.

Figure~1 shows a schematic representation of our experimental setup named X-ray laser-enhanced attosecond pulse generation (XLEAP). 
The energy distribution of the electron beam is modulated by the resonant interaction with a high-power infrared pulse in a long-period undulator~(or wiggler)~\cite{zholents_method_2005,ESASE,hemsing2014beam}. 
This modulation is converted into one or more high-current ($\sim 10$~kA) spikes by a magnetic chicane. 
The spikes are subsequently used in the undulator to generate short X-ray pulses. 
This bunch compression method effectively broadens the XFEL bandwidth and allows the generation of sub-fs pulses in the soft X-ray spectral region. 
In our experiment, rather than using an external infrared laser as originally proposed by Zholents \cite{zholents_method_2005}, we employ the coherent infrared radiation emitted by the tail of the electron beam in the wiggler to modulate the core of the electron beam~\cite{MacArthurSELFMOD,JamesSELFMOD}. 
This method results in a phase-stable, quasi-single-cycle modulation, and naturally produces a single high-current spike that can generate an isolated attosecond pulse. 
Figures~1~(a-d) show the measured initial current profile and the evolution of the phase-space of the core of the electron bunch during the three stages of ESASE compression. 

After separating the broad bandwidth X-ray pulses from the spent electron bunch, the X-ray pulses are focused and temporally overlapped with a circularly polarized, 1.3~$\mu$m, infrared~(IR) laser field in a velocity map imaging~(VMI) spectrometer~\cite{li_co-axial_2018}. 
Photoelectrons ionized by the X-ray pulse receive a ``kick'' proportional to the vector potential of the IR laser pulse at the time of ionization~\cite{kienberger_atomic_2004}. 
Through the interaction of the ionized electron with the dressing IR-laser field, the temporal properties of the X-ray pulse are mapped onto the final momentum distribution of the emitted photoelectrons~\cite{kazansky_interference_2016,kazansky_angular_2017,li_characterizing_2018}.
This technique was originally called the ``attosecond streak camera,'' and is routinely used to measure the temporal profile of isolated attosecond pulses from HHG sources~\cite{itatani_attosecond_2002}.
In contrast to measurements done with HHG sources, in this work we are able to diagnose the single-shot pulse profile, 
rather than 
an average pulse shape.
Moreover, the shot-to-shot fluctuations~(or jitter) in the relative arrival time between the X-ray and optical field present at an FEL facility~\cite{glownia_time-resolved_2010}  makes single-shot measurements unavoidable. 
This single shot measurement scheme was originally demonstrated at LCLS by Hartmann~\textit{et al.}, who recovered the ``time-energy structure'' of SASE pulses produced by LCLS~\cite{hartmann_attosecond_2018}. 
We have adapted this technique to measure the sub-femtosecond structure of the X-ray pulses produced by XLEAP.

\section*{Results}
Figure~2~(a) shows a single-shot measurement of the ``streaked'' photoelectron momentum distribution, 
which we use to reconstruct the full temporal profile of the X-ray pulse~\cite{li_characterizing_2018}.
The raw data is filtered and down-sampled~(Fig.~2~(b)) before being fed into the reconstruction algorithm, which returns a pulse profile and corresponding photoelectron distribution~(Fig.~2~(c)).
The robustness of this algorithm has been tested at length in Ref.~\cite{li_characterizing_2018}, and is detailed in the supplemental material. 
Figure~2 also shows representative temporal profiles retrieved from the reconstruction at photon energies of  905~eV~(panel~d) and 570~eV~(panel~e).
Figures~2~(f)~and~(g) show the distribution of pulse widths~(full width at half maximum of the intensity profile) retrieved from two large data sets at these photon energies. 
The data shows that the XLEAP setup generates sub-femtosecond X-ray pulses, and we find a median duration of 284~as FWHM (476~as) at 905~eV (570~eV). 
The pulse duration fluctuates on a shot-to-shot basis and half of the single-shot measurements fall within a 106~as (166~as) window  at 905~eV (570~eV). 
This amount of fluctuations is consistent with numerical simulations of ESASE FEL operation (see e.g. \cite{Zhang2color}). 
The estimated uncertainty on the single-shot pulse duration is between 10\% and 30\% of the measured duration depending on the pulse energy and the amplitude of the streaking laser field (a discussion on the experimental uncertainty of the measurement can be found in the supplemental materials).  
The median pulse energy is 10~$\mu$J at 905~eV and 25~$\mu$J at 570~eV. 
However, due to the intrinsic fluctuations of SASE FELs~\cite{PellegriniSpectrum} we observe pulses well above the mean value (up to 250~$\mu$J for 570~eV, corresponding to a peak power in the hundreds of GW).
We note that for the 570~eV dataset we were only able to obtain converging reconstructions for pulse energies higher than 130~$\mu$J, corresponding to the top 8\%. 
However, since the data at both energies does not show a significant correlation between pulse energy and duration~(Figure~2 panels~(b)~and~(e)) we believe that the average pulse duration from this sample is representative of the entire data set.

In a separate set of experiments we measured single-shot X-ray spectra with a grating spectrometer.
Figure~3 shows a range of single-shot X-ray spectra recorded around 650~eV and 905~eV, and the distribution of the measured bandwidth~(FWHM).
The median FWHM bandwidth is 7.5 eV and 5 eV for the 905~eV  and 650~eV datasets respectively.
The Fourier transform limited~(FTL) duration for a bandwidth of 7.5~eV~(5~eV) is 240~as~(365~as). 
The average pulse duration recovered from our reconstruction at similar energies is within  a factor of 2 of the FTL value. 
This discrepancy is due to the beam-energy chirp introduced by longitudinal space-charge forces within the high-current ESASE spike~\cite{ESASEding}. 
This results in a residual chirp in the emitted X-rays, which is reproduced in the reconstruction (see Supplemental Materials). 
Ripples in the spectral intensity are visible in the 650 eV spectra and are due to interference with satellite pulses. 
The pulse energies of these side pulses can be inferred from the single-shot spectra and are typically less than 0.3\% of the main pulse for 650~eV and negligible for 905~eV.


\section*{Considerations for Pump/Probe Spectroscopy}
To put the results of our work in context, we detail the development of isolated attosecond pulse sources in Figure~4, where we compare the measured pulse energy from existing attosecond light sources with the requisite flux to saturate the ionization of $1s$ electrons in various atomic systems.
The saturation level serves as a coarse approximation to the energy required for a pump/probe experiment, and sources within two orders of magnitude of saturation are likely useful for pump/probe studies. 
The pulse energy produced by HHG sources decays very rapidly with the photon energy and is several orders of magnitude below the threshold for non-linear interaction in the soft X-ray range~($E > 280$~eV).
Conversely our method can produce isolated attosecond pulses with tens of $\mu$J of pulse energy, increasing the available pulse energy at soft X-ray wavelengths by six orders of magnitude, and reaching intensities sufficient for attosecond pump/attosecond probe experiments. 
Note that Fig.~4 reports the pulse energy measured for the experiments shown in Figs.~2~and~3, as well as other experiments using the XLEAP setup at different photon energies. 
The highest observed median pulse energy is $\sim50~\mu$J.



In addition to high single pulse photon flux, the application of this technique to attosecond pump/attosecond probe experiments requires the generation of pairs of synchronized pulses.
Ideally these pulses could have different photon energies allowing for excitation at one atomic site in a molecular system to be probed at another~\cite{mukamel_multidimensional_2013}.
To this end, ESASE can be easily adapted to generate pairs of pulses of different colors using the split undulator method~\cite{Lutman2C,SACLA2C,marinelli_multicolor_2013}.
In this scheme the LCLS undulator is divided in two parts separated by a magnetic chicane, as shown at the top of Fig.~5. 
The ESASE current spike is used to generate two X-ray pulses of different energies in the two undulators. The magnetic chicane delays the electrons with respect to the X-rays, thus introducing a controllable delay between the first and second X-ray pulses. 

Figure~5 shows the results of such a double-pulse ESASE experiment at LCLS. 
Two pulses with an average pulse energy of $6~\mu$J each and an energy separation of 15~eV were generated~(see Fig.~5~b).
The timing jitter between the two pulses was not measured but numerical simulations indicate that it is shorter than the individual pulse duration. 
We note that the energy separation range in our experiment is limited by the tuning range of the LCLS undulator (roughly 3\% of the photon energy~\cite{Lutman2C}), but this scheme could be used with variable gap undulators and allow fully independent tuning of the two colors. 
This will be possible with the upcoming LCLS-II upgrade, enabling continuous tuning between 250~eV and 1200~eV~\cite{schoenlein_new_2015}. 
The temporal separation can be varied from a minimum of 2~fs up to a maximum of roughly 50~fs.
Smaller delays could be accessed with a gain-modulation scheme \cite{marinelli_multicolor_2013}. 
Improved two-color operation with higher peak power and delay control through overlap could be achieved with a modest upgrade of the XLEAP setup~\cite{Zhang2color}.


Using the split-undulator scheme shown in Fig.~5~a, one can also generate two pulses of the same photon energy and with mutual phase stability. 
Unlike the case of two different colors, where the two pulses are seeded by noise at different frequencies and are uncorrelated, in this case the beam microbunching that generates the first pulse is re-used to generate a second pulse and the two are phase-locked. 
Figure~5c shows the measured spectra under these conditions. 
The spectra exhibit stable and repeatable fringes, which implies that the phase between the two pulses is stable to better than the X-ray wavelength. 
From the variation in the spectral fringes we can infer a phase jitter of 0.77~rad, or 0.5~as between the pulses.
In this case the delay can be varied from 0~fs to roughly 5~fs. 
Beyond this value the delay chicane will destroy the X-ray microbunching and hence the phase stability of the pulses.

\section*{Summary and conclusions}
We have demonstrated tunable sub-femtosecond X-ray pulses with tens of gigawatts of peak power using a free-electron laser. 
The pulses were generated by an electron bunch modulated by interaction with a high-power infrared light pulse and compressed in a small magnetic chicane. 
To diagnose the temporal structure of these pulses we used an attosecond streak camera and measured a median pulse duration of 284~as~(476~as) at 905~eV~(570~eV).
With an eye towards pump/probe experiments, pairs of sub-fs pulses were demonstrated using a split undulator technique, showing control of the delay and energy separation, but with a reduced peak power compared to the single-pulse case. 

These pulses have pulse energies six orders of magnitude higher than what can be achieved with HHG based sources in the same wavelength range. The measured peak power is in the tens to hundreds of gigawatts.
Such a marked increase in pulse energy will enable a suite of non-linear spectroscopy methods such as attosecond-pump/attosecond-probe experiments~\cite{leone_what_2014,schweigert_probing_2007} and four-wave mixing protocols~\cite{mukamel_multidimensional_2013}. 
Moreover, the achieved photon flux is will enable single-shot X-ray imaging at the attosecond timescale. Finally, the XLEAP setup is based on a passive modulator and it is naturally scalable to the MHz-repetition rate envisioned for the next generation of X-ray free-electron lasers \cite{schoenlein_new_2015,altarelli2011european}.






\clearpage


\resizebox{0.95\columnwidth}{!}{\includegraphics{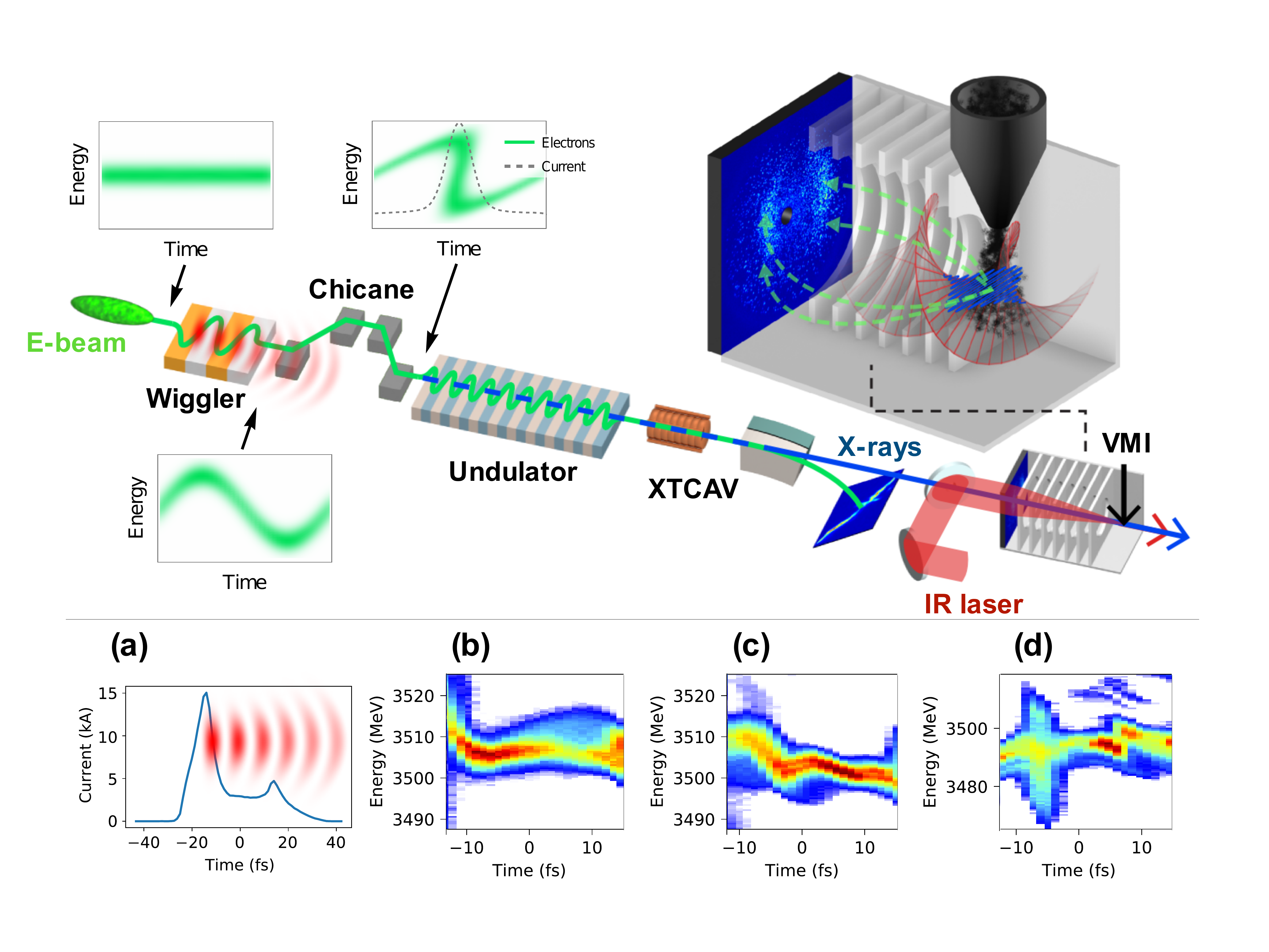}} \\
\noindent {\bf Fig. 1.} {\bf Top: schematic representation of the experiment.} The electron beam travels through a long period (32 cm) wiggler and develops a single-cycle energy modulation. The energy modulation is turned into a density spike by a magnetic chicane and sent to the LCLS undulator to generate sub-fs X-ray pulses. After the undulator the relativistic electrons are separated from the X-rays and sent to a transverse cavity (labeled XTCAV) used for longitudinal measurements of the beam.
The X-rays are overlapped with a circularly polarized infrared laser (labeled IR laser) and interact with a gas-jet to generate photoelectrons. The extracted photoelectrons are streaked by the laser and detected with a velocity map imaging~(VMI) spectrometer. The momentum distribution of the electrons is used to reconstruct the pulse profile in the time domain. 
 {\bf Bottom: measurements of the ESASE modulation process.} (a)~the measured current profile of the electron bunch generated by the accelerator. The tail of the bunch has a high-current horn that generates a high-power infrared pulse used for the ESASE compression. (b),~(c)~and~(d)~Show the longitudinal phase-space of the core of the electron bunch in three different conditions: (b) with no wiggler and no chicane we measure the electron distribution generated by the accelerator; (c)  after inserting the wiggler we observe a single-cycle energy modulation generated by the interaction between electrons and radiation; (d) after turning on the chicane the modulation is turned into a high current spike at t = -5 fs.

\pagebreak    
\resizebox{0.95\columnwidth}{!}{\includegraphics{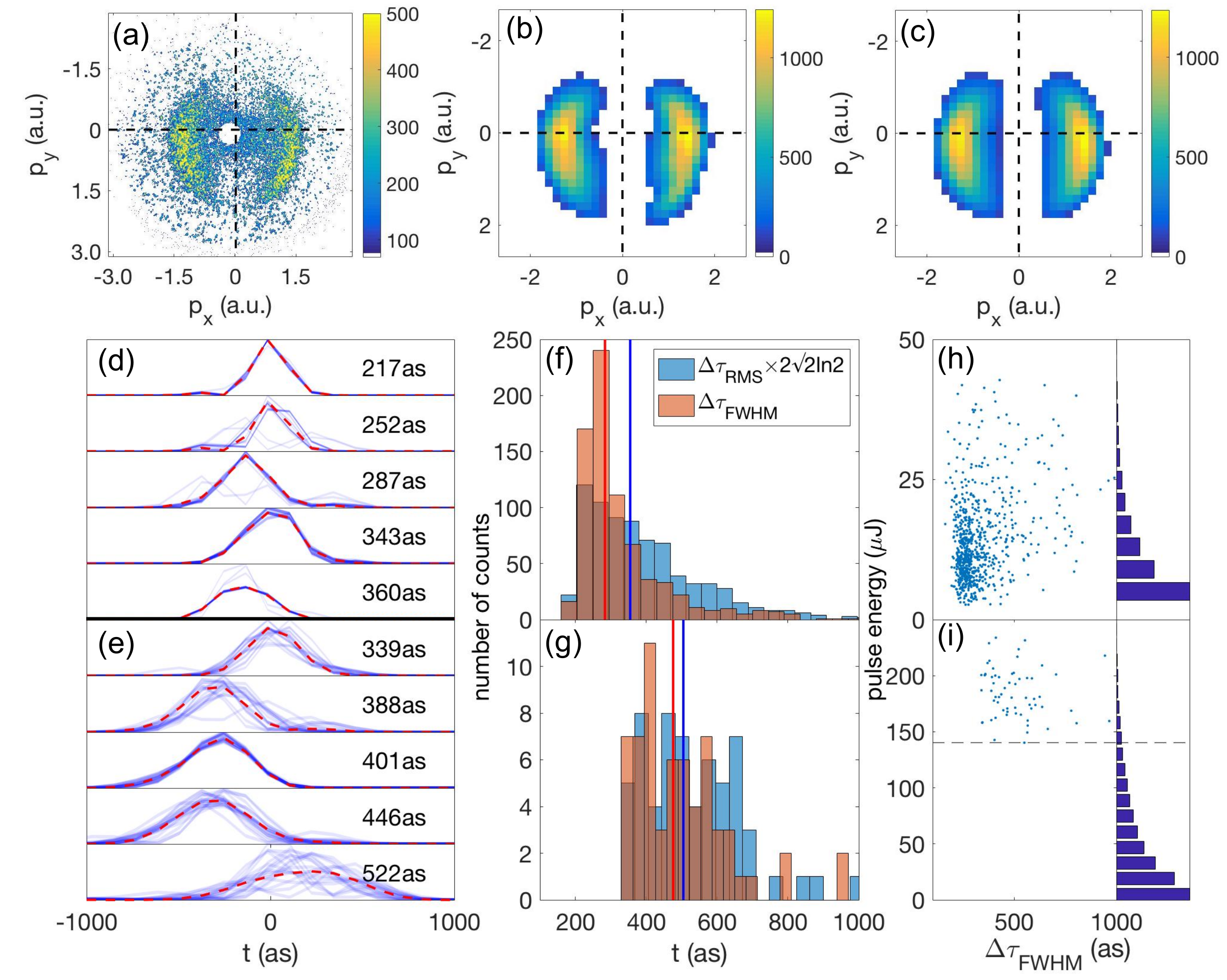}} \\
\noindent {\bf Fig. 2.} 
{\bf Results of the angular streaking measurement.}
(a,b,c): measured and reconstructed streaked photoelectron distribution from a single X-ray pulse.
Our reconstruction algorithm reads the photoelectron momentum distribution (a), downsamples the data (b) and fits it to simulated streaked spectra calculated from a complete basis set (c). 
(d,e) Representative pulse reconstruction at 905~eV (d) and 570~eV (e). The shaded blue lines represent the solutions found from running the reconstruction algorithm initiated by different random seeds, and the red lines represent the most probable solution (see Supplementary Material for details). The labeled number is the averaged $\Delta\tau_{\rm{FWHM}}$ over the different solutions.
(f,g): distribution of retrieved X-ray pulse durations for 905~$\rm{eV}$~(f) and 570~$\rm{eV}$~(g). The red and blue vertical lines correspond to the median of $\Delta\tau_{\rm{FWHM}}$ and $\Delta\tau_{\rm{RMS}}\times2\sqrt{2\rm{ln}2}$ respectively. For 905~$\rm{eV}$ data, they are 284~$\rm{as}$ and 355~$\rm{as}$. For 570~$\rm{eV}$ data, they are 476~$\rm{as}$ and 505~$\rm{as}$. 
(h,i): scatter plot of pulse energy as a function of pulse duration for the reconstructed shots and a histogram of the pulse energy for the entire data set for 905~$\rm{eV}$ (h) and 570~$\rm{eV}$ (i).

 
\pagebreak    
\resizebox{0.65\columnwidth}{!}{\includegraphics{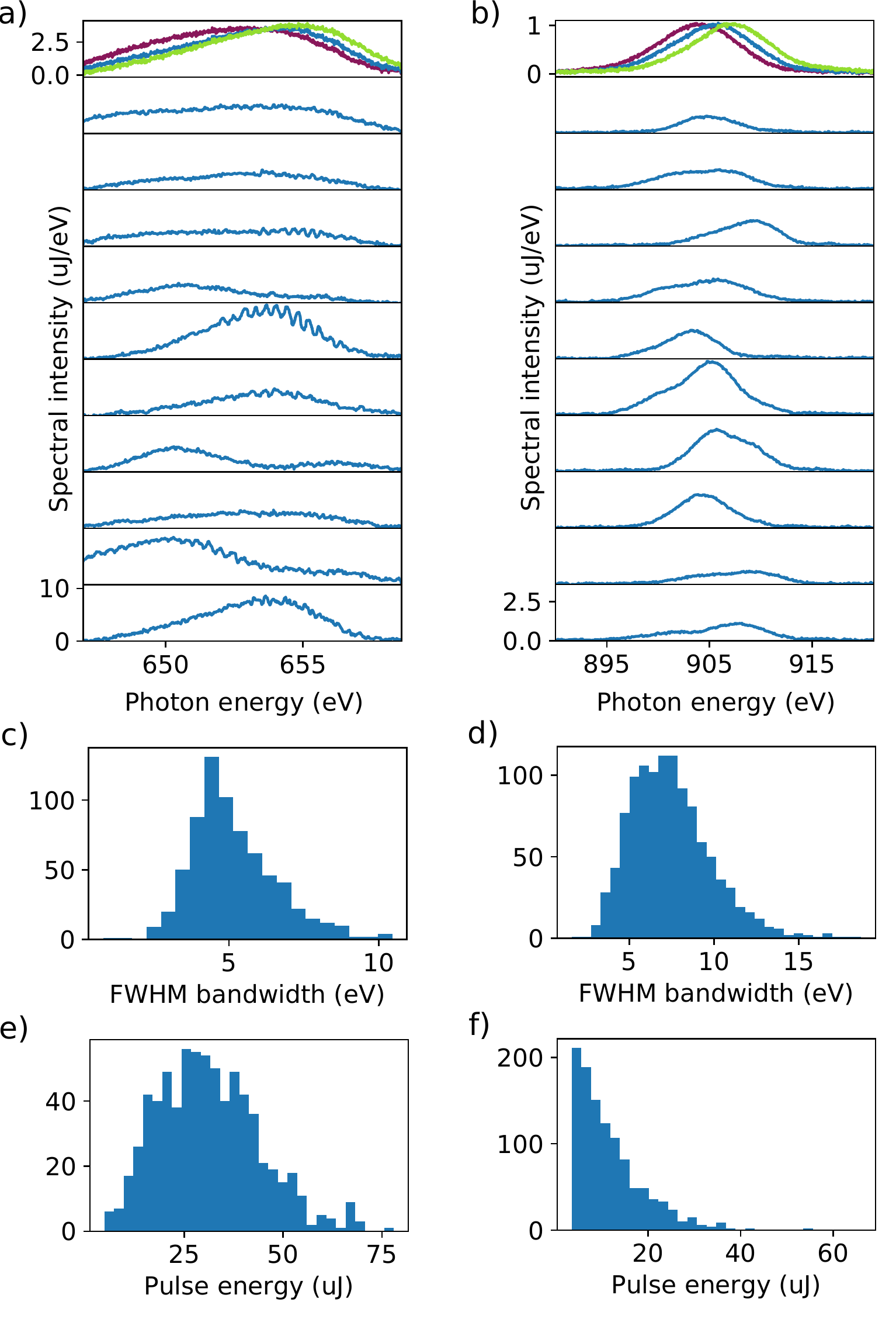}} \\
\noindent {\bf Fig. 3.} 
(a,b) Spectra of the attosecond X-ray pulses measured with a grating spectrometer at two different electron beam energies ((a): 3782.1 MeV; (b): 4500.3 MeV). The top figures show average spectra at slightly different electron beam energies (in steps of 2.7~MeV for 650~$\rm{eV}$ and 3.0~$\rm{MeV}$ steps for 905~$\rm{eV}$), while the remaining spectra show single-shot measurements at the energy of the center curves shown in the top plots. (c,d):  histogram of the distribution of FWHM bandwidths. Panels (e,f) show a histogram of the distribution of pulse energies.

\pagebreak
\resizebox{0.75\columnwidth}{!}{\includegraphics{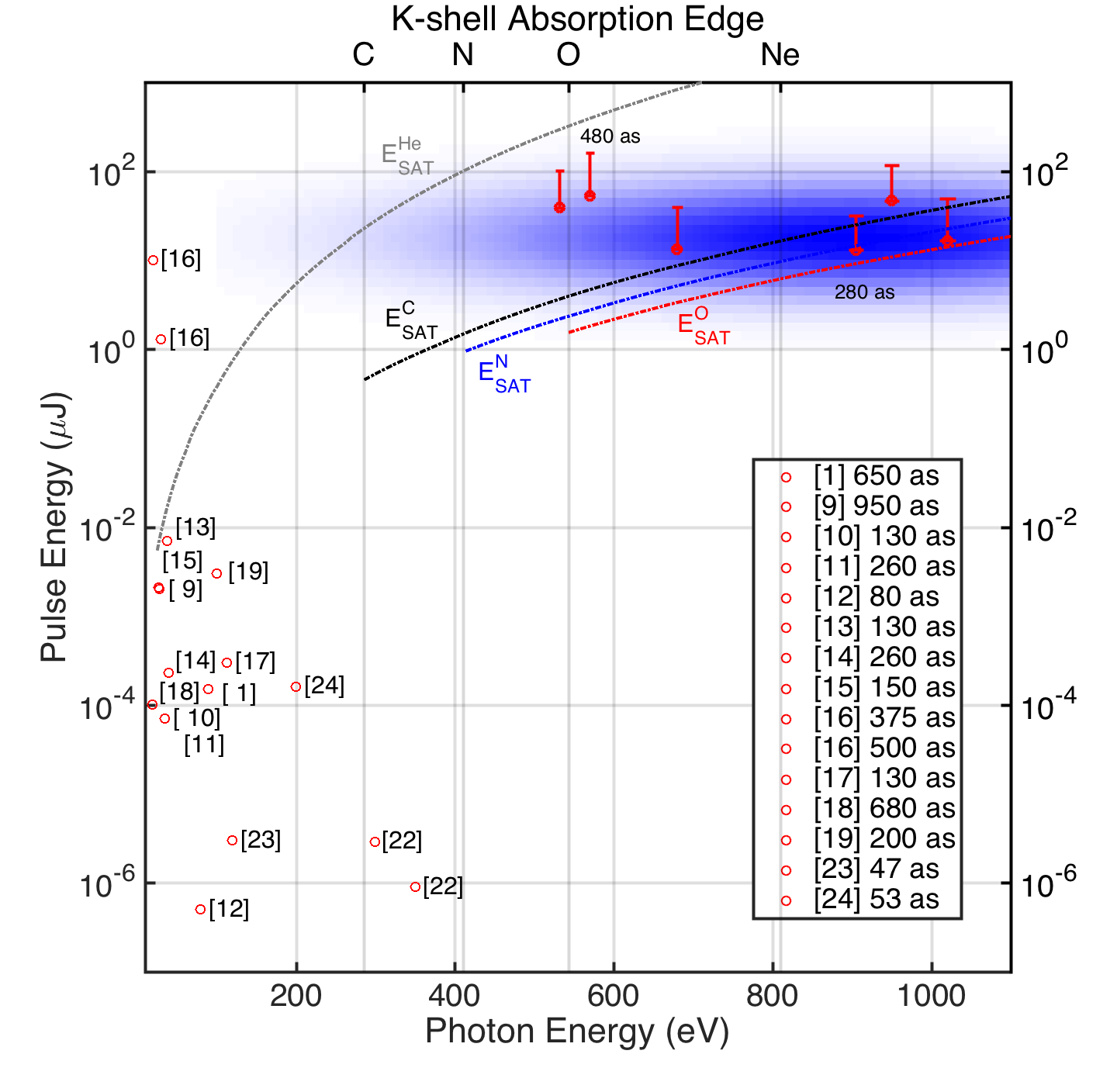}} \\
\noindent {\bf Fig. 4.} Survey of published isolated attosecond pulse sources~\cite{hentschel_attosecond_2001,sekikawa_nonlinear_2004,sansone_isolated_2006,sola_controlling_2006,goulielmakis_single-cycle_2008,mashiko_double_2008,feng_generation_2009,ferrari_high-energy_2010,takahashi_attosecond_2013,ossiander_attosecond_2017,barillot_towards_2017,bergues_tabletop_2018,teichmann_0.5-kev_2016,gaumnitz_streaking_2017,li_53-attosecond_2017} extending into the soft X-ray domain~(red open circles), along with the results demonstrated in this work for a number of different photon energies~(red filled circles). 
The filled circle shows the average pulse energy recorded during the experiment, the error bar extends from the central energy and includes up to 90$\%$ of the recorded pulse energies.   
All previous results were obtained via strong-field driven high harmonic generation with near-infrared and mid-infrared laser fields, while our results are obtained using a free-electron laser~(FEL) source. 
As a first-order estimate of the propensity for nonlinear science (pump/probe spectroscopy, etc.) we show the pulse energy required to saturate $1s$ ionization of carbon~(dash-dot, black), nitrogen~(dash-dot, blue), oxygen~(dash-dot, red), and helium~(dash-dot, gray), assuming a $1~\mu$m$^2$ focal spot size, as a function of X-ray photon energy.
Sources within 2 orders of magnitude of the line are likely sources for pump-probe studies.
The table in the bottom right corner gives the published pulse duration for the previous measurements. 
The shaded blue area shows the operational range predicted for LCLS-II.

\pagebreak
\resizebox{0.9\columnwidth}{!}{\includegraphics{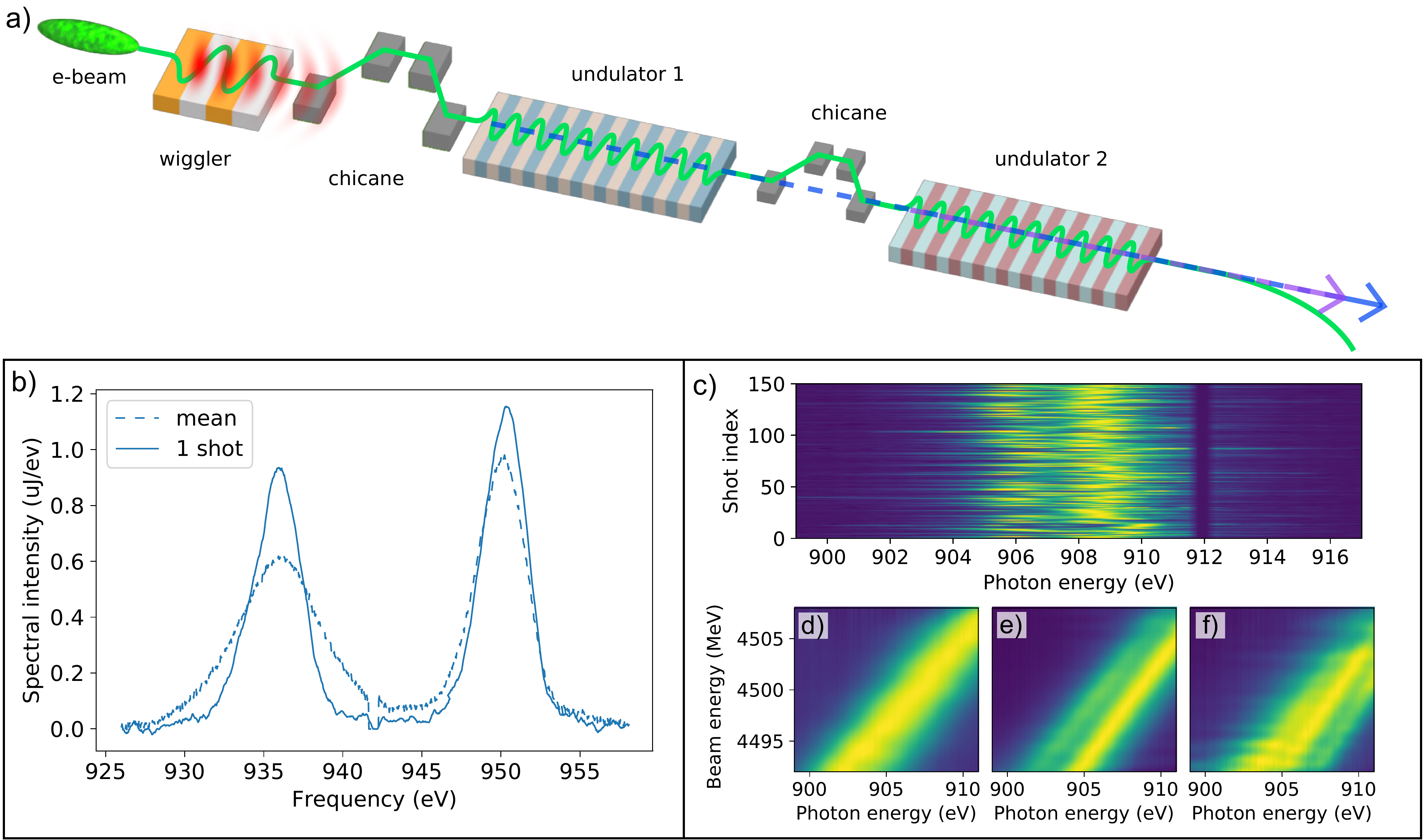}} \\
\noindent {\bf Fig. 5.} (a): Schematic representation of the double pulse generation experiment. The electron beam is modulated and compressed in the XLEAP beamline and sent to the LCLS undulator. The undulator is divided in two parts separated by a magnetic chicane. Each half of the undulator is used to generate an X-ray pulse with different pulse energies, and the chicane introduces a variable delay between the pulses.
(b): Single shot and average two-color spectra measured with a grating spectrometer. (c): Single-shot measurements of the spectrum of the pulse pair with 1~fs delay. The repeatable spectral fringe demonstrate phase stability between the pulses. (d), (e), (f): Averaged phase-stable double shot spectra as a function of photon energy and electron beam energy for a nominal chicane delay of 0~$\rm{fs}$ (single pulse)(d), 0.5~$\rm{fs}$ (e) and 3~$\rm{fs}$ (f).

\pagebreak
\section*{Supplementary materials}
Materials and Methods\\
Supplementary Text\\
Figs. S1 to S12\\
References \textit{(65-78)}

\bibliographystyle{science.bst}
\bibliography{referencesJPC.bib,biblio.bib}

\section*{Acknowledgements}
The authors would like to acknowledge Tais Gorkhover, Christoph Bostedt, Jon Marangos, Claudio Pellegrini, Adrian Cavalieri, Nora Berrah, Linda Young, Lou DiMauro, Heinz-Dieter Nuhn, Gabriel Marcus, and Tim Maxwell for useful discussions and suggestions. We would also like to acknowledge Michael Merritt, Oliver Schmidt, Nikita Strelnikov, Isaac Vasserman for their assistance in designing, constructing and installing the XLEAP wiggler.
We also acknowledge the SLAC Accelerator Operations group, and the Mechanical and Electrical engineering divisions of the SLAC Accelerator Directorate, especially Gene Kraft, Manny Carrasco, Antonio Cedillos, Kristi Luchini and Jeremy Mock for their invaluable support.

This work was supported by U.S. Department of Energy Contracts No. DE-AC02-76SF00515, DOE-BES Accelerator and detector research program Field Work Proposal 100317, DOE-BES, Chemical Sciences, Geosciences, and Biosciences Division, and  Department of Energy, Laboratory Directed Research and Development program at SLAC National Accelerator Laboratory, under contract DE-AC02-76SF00515. WH acknowledges financial support by the BACATEC programme. P.R. and M.F.K. acknowledge additional support by the DFG via KL-1439/10, and the Max Planck Society. G. H. acknowledges the Deutsche Forschungsgemeinschaft (DFG, German Research Foundation) – Projektnummer 328961117 – SFB 1319 ELCH.

\end{document}



\baselineskip24pt


\maketitle 
\pagebreak
\tableofcontents

\renewcommand{\thefigure}{S\arabic{figure}}
\setcounter{figure}{0}

\section{Materials and Methods}
\subsection{Beam Dynamics and FEL Configuration}
The x-ray free-electron laser (XFEL) at the Linac Coherent Light Source (LCLS) is composed of a high-brightness linear accelerator (linac) and a magnetic undulator. The XLEAP beamline is composed of a long-period wiggler and a magnetic chicane prior to the undulator section. 
The accelerator and undulator/wiggler parameters used in this experiment are summarized in Tab. 1. 
\newline
\\
\begin{tabular} {| l | p{5cm} |}
\hline
\multicolumn{2}{|l|}{Accelerator and Undulator Parameters} \\ \hline
{Beam Energy} &  3-5 GeV \\ \hline
{Repetition rate} &  120 Hz \\ \hline
{Normalized Emittance} &  0.4 $\mu m$ \\\hline
{Peak Current Before ESASE Compression} &  2.5 - 3.5 kA  \\ \hline
{Peak Current After ESASE Compression (estimated)} & $\sim$ 10  kA \\ \hline
{XLEAP Wiggler Period $\lambda_w $} & 32 cm  \\ \hline
{XLEAP Wiggler Parameter $K_w$} & 0 - 52 \\ \hline
{XLEAP Chicane Longitunal Dispersion $R_{56}$} & 0 - 0.9 mm \\ \hline
{Undulator Period $\lambda_u$} & 3 cm  \\ \hline
{Undulator Parameter $K_u$} & 3.45 - 3.51 \\ \hline
\end{tabular}
\newline
\newline
The radiation wavelength generated by the XFEL is given by the well known resonant formula \cite{BPN}:
\begin{equation}
    \lambda_r = \lambda_u \frac{1+\frac{K_u^2}{2}}{2\gamma^2}
\end{equation}
where $\gamma$ is the Lorentz factor of the electron beam (between 6000 and 10000 for this experiment). Similarly, the radiation wavelength  generated in the wiggler (and therefore the electron beam modulation wavelength) is given by
\begin{equation}
    \lambda_{IR} = \lambda_w \frac{1+\frac{K_w^2}{2}}{2\gamma^2}.
\end{equation}

The XLEAP beamline was designed to provide a modulation wavelength in the range between 2 and 4 $\mu m$ for a beam energy between 3 and 5 GeV. At these beam energies the LCLS undulator produces photons with an energy between $\sim$400 and $\sim$1200 eV. The amplitude of the observed energy modulation is typically in the few MeV range and can be fully compressed with the available dispersion of the XLEAP chicane.
The self-modulation process has been described in detail in \cite{MacArthurSELFMOD}.

The high-current spike generated by the XLEAP modulator generates strong longitudinal space-charge forces which result in a strong time-energy correlation on the electrons within the spike \cite{ESASEding}. 
The longitudinal space-charge force for a highly relativistic beam in an undulator is proportional to the derivative of the longitudinal current profile:
\begin{equation}
    E_z (s) = -\frac{Z_0 I'(s) (1+\frac{K^2}{2})}{4\pi \gamma^2} \left(2 \log\left(\frac{\gamma \sigma_z}{r_b \sqrt{1+\frac{K^2}{2}}}\right) \right)
\end{equation}
where $I(s)$ is the beam current as a function of the longitudinal coordinate $s$, $\sigma_z$ is the length of the current spike, and $r_b$ is the beam radius.
For a Gaussian-like spike such as the one generated by an ESASE modulator, the correlated energy spread (or chirp) is predominantly linear but has a non-vanishing nonlinear component.
The observed final energy distribution has a full width in the range between 30 and 40 MeV, over a spike duration of order 1 fs. 
The bandwidth of this chirp is much larger than the acceptance of the FEL. 

To preserve the FEL gain with such a large energy spread the undulator parameter $K$ has to be varied (tapered) along the undulator length so that the resonant condition is maintained as the photons slip ahead of the electrons and interact with electrons of a different energy \cite{SaldinChirpTap, PanosAtto,ESASEding}. This chirp-taper matching condition can be expressed as:
\begin{equation}
    \frac{K}{1+K^2}\frac{dK_u}{dz}\lambda_w = \frac{d\gamma}{ds} \lambda_r,
    \label{chirptap}
\end{equation}
where 
$z$ is the position along the undulator beamline.
The chirp-taper matching condition  can only be verified for a specific value of the linear chirp $d\gamma /ds$, meaning that if we optimize the undulator to lase in the ESASE spike, the rest of the bunch will be mismatched in energy, therefore suppressing the background radiation outside of the main pulse.
Furthermore it can be shown that the cubic chirp introduced by space-charge contributes to shortening the pulse duration when the taper is optimized to compensate the linear chirp \cite{PanosAtto}. 

\begin{figure}
    \centering
    \resizebox{0.65\columnwidth}{!}{\includegraphics{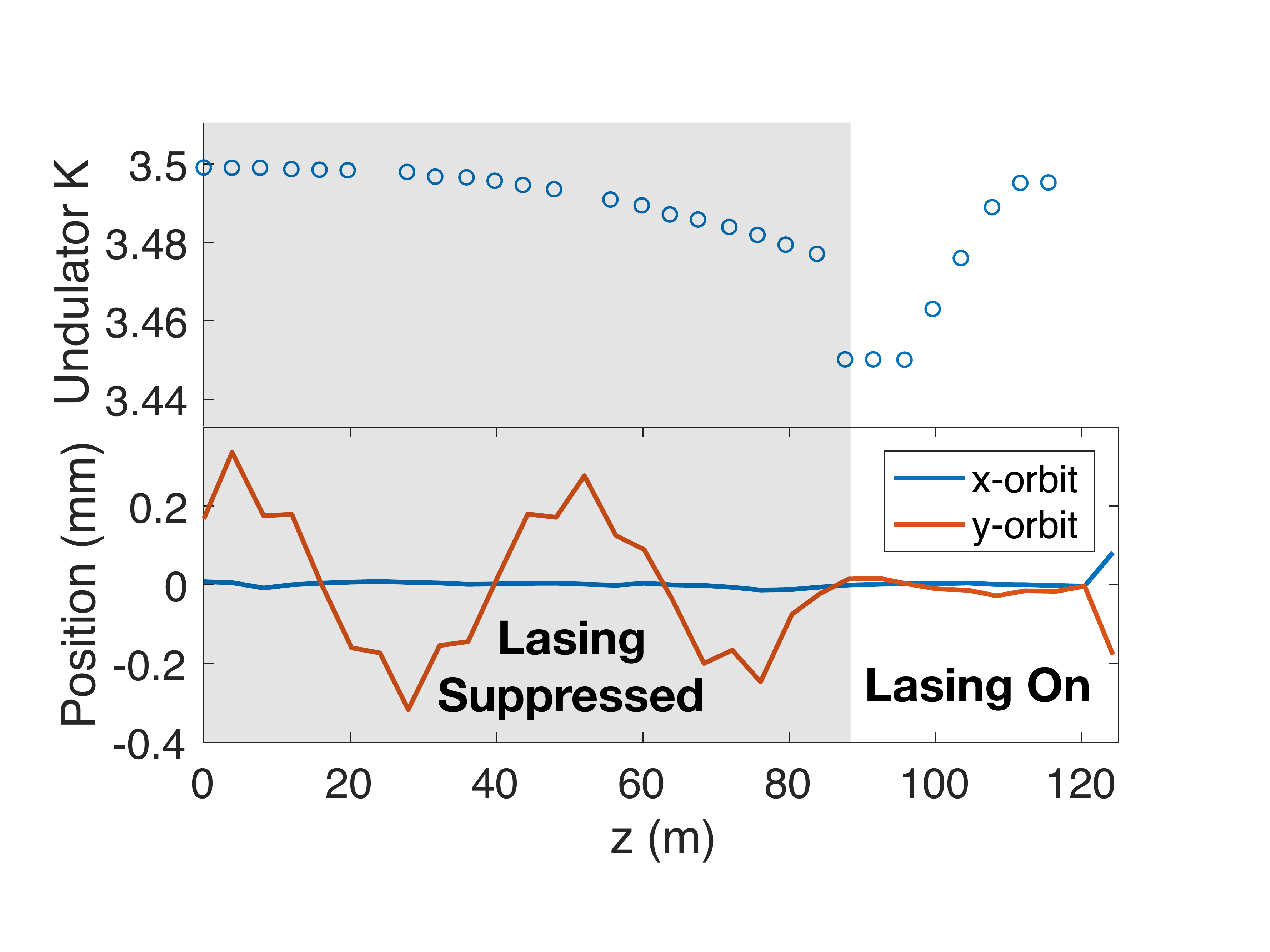}}
    \caption{Top: typical undulator configuration for an ESASE experiment. The FEL gain is suppressed in the first 24 undulators by a large oscillation in the vertical trajectory of the electron beam. Bottom: measured vertical and horizontal trajectory through the undulators.}
    \label{fig::UndSetup}
\end{figure}

To take advantage of these two effects we maximized the chirp introduced by space-charge by only allowing the electron beam to lase in the last few sections of our undulator beamline, and suppressing the FEL gain in the first part of the undulator by introducing a large transverse oscillation in the electron beam trajectory.
Figure~\ref{fig::UndSetup} shows an example of a typical undulator configuration for an ESASE experiment.
The last 7 undulators are used for lasing, while the remaining 24 are used to maximize the energy chirp introduced by space-charge. The electron beam trajectory is also plotted, showing a large vertical oscillation in the first 24 undulators to suppress the FEL gain.

The space-charge induced energy-spread can be seen in Fig. 1-d of the main text, which shows the measured time-energy distribution of the full electron bunch. The ESASE current spike is seen as a large vertical stripe, i.e. a short region of the bunch with a large energy spread. The resolution of our diagnostic is not enough to resolve the sub-fs structure of the phase-space, and therefore the chirp of the ESASE spike itself appears as a vertical stripe with no apparent time-energy correlation.

\subsubsection{Longitudinal Phase-Space Measurements}
\begin{figure}
    \centering
    \resizebox{0.95\columnwidth}{!}{\includegraphics{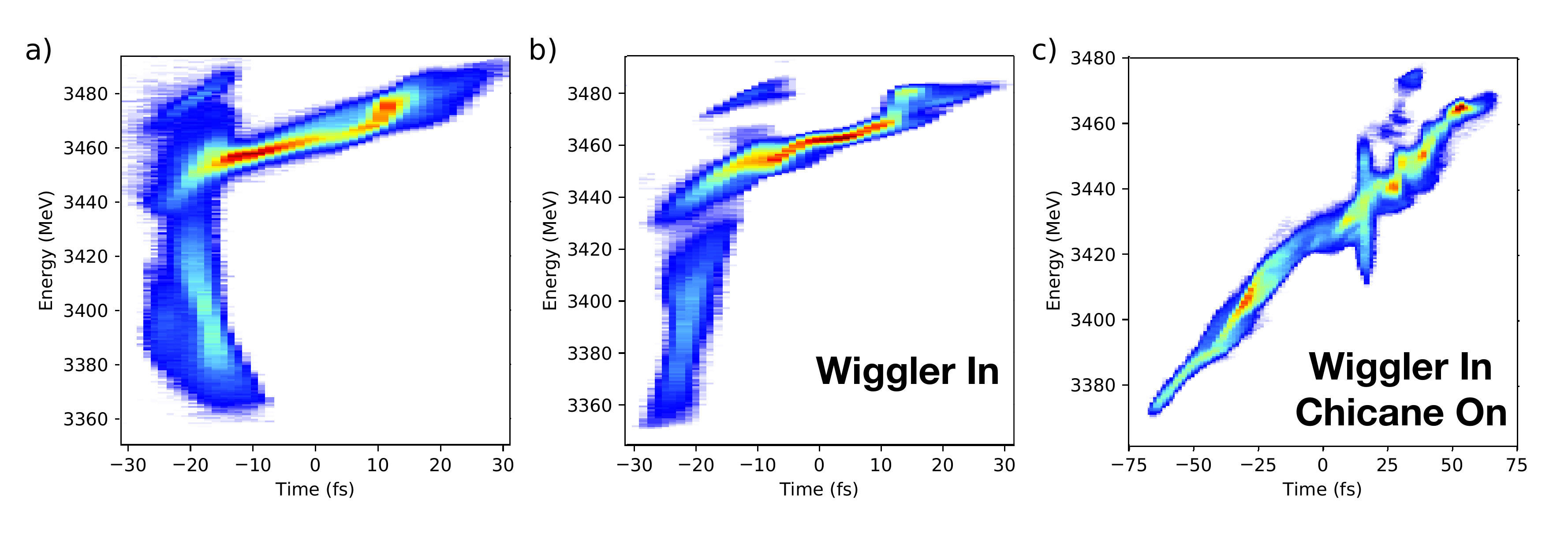}}
    \caption{Measured longitudinal phase space for three configurations: (a) wiggler out and chicane off, (b) wiggler in and chicane off, (c) wiggler in and chicane on. The data in Figs. 1-a,b,c,d was processed from these three measurements.}
    \label{fig::XTCAV}
\end{figure}
The time-energy distribution of the electrons (the longitudinal phase-space) can be measured on a shot-by-shot basis in a non-destructive fashion (i.e. without intercepting or otherwise affecting the x-rays). The measurements are performed with an x-band transverse cavity placed in the dump line of the LCLS~\cite{XTCAV}.

Figures 1-b,c,d of the main article are a zoomed-in version of the full measured electron beam phase-space. Figure~\ref{fig::XTCAV} reports the entire longitudinal phase-space measurement for the same shots. To better represent the phase-space in the XLEAP beamline we have numerically removed the energy variation introduced by the wakefield of the undulator vacuum chamber from Figures 1-b,c,d. The x-band transverse cavity diagnostic does not have enough temporal resolution to measure the duration of the ESASE spike in Fig.~1~(d), but numerical simulations of this technique suggest that the peak current is close to 10~kA, with a spike duration of roughly 1~fs.

\subsection{Angular Streaking Setup}
\begin{figure}
    \centering
    \resizebox{0.65\columnwidth}{!}{
    \includegraphics{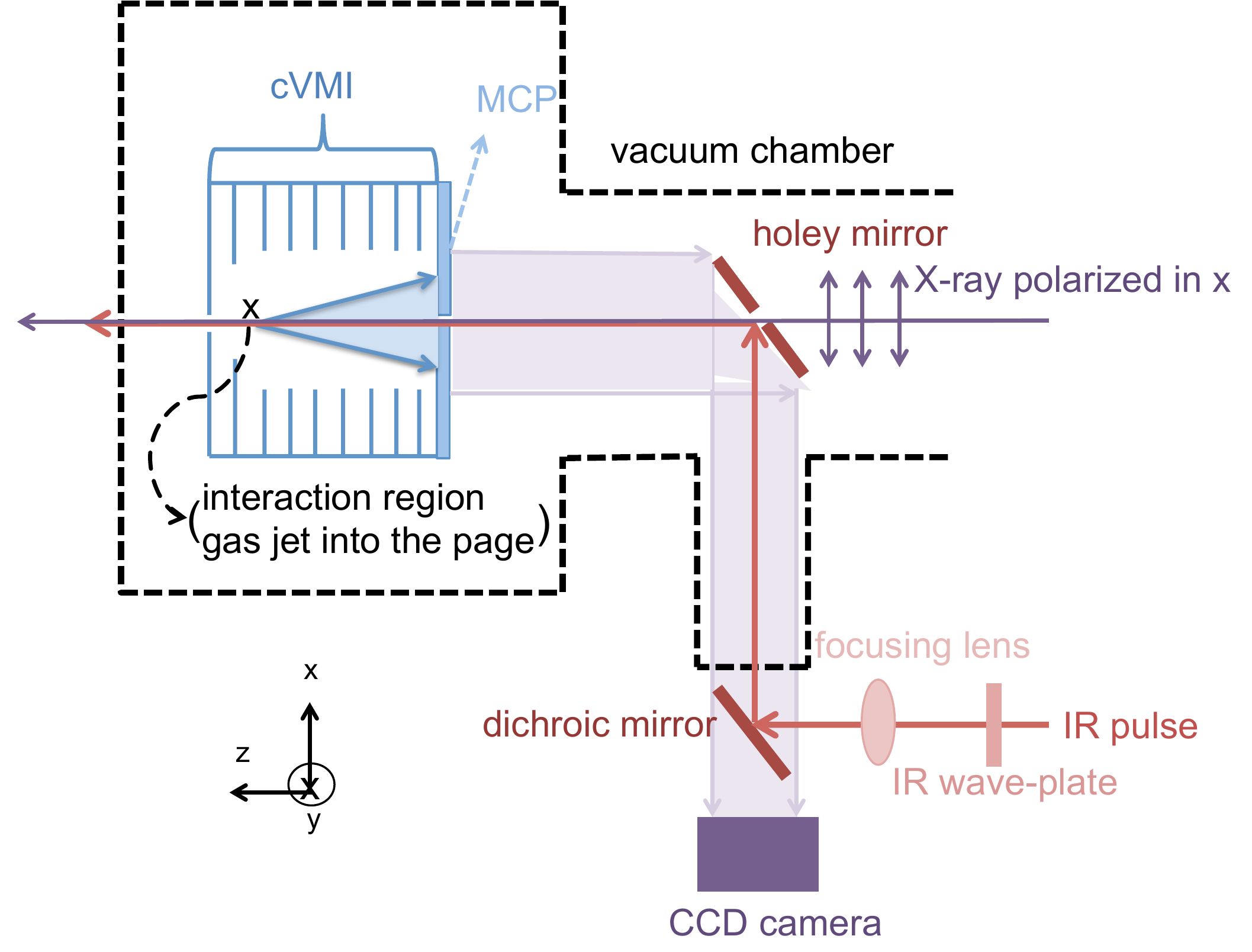}}
    \caption{Experimental geometry of the co-axial velocity map imaging~(c-VMI) apparatus used in the angular streaking setup. Linearly polarized x-rays pass through a 2 mm hole in a silver coated mirror before coming to a focus in the interaction region of the c–VMI spectrometer. Photoelectrons are extracted by the electrostatic field of the spectrometer, in the direction opposite of the x-ray propagation. The phosphor screen of the charged particle detector is imaged via the holey silver mirror, through a dichroic mirror used to introduce an IR laser pulse. The polarization of the IR pulse is controlled via a quarter wave-plate upstream of the dichroic mirror. The target gas is introduced via a skimmed molecular beam oriented in the $y$-direction.}
    \label{fig::ExpSetup}
\end{figure}

Our experiment was performed at the Atomic, Molecular, and Optical physics~(AMO) beamline of the Linac Coherent Light Source (LCLS).
The streaking laser pulse is derived from a 120~Hz titanium-doped sapphire laser system synchronized to the accelerator.
Ten mJ, 800~nm laser pulses are compressed to $\sim40$~fs, and the compressed pulse is used to pump an optical parametric amplifier~(TOPAS-HE, Light Conversion) which produces 500~$\mu$J pulses at a wavelength of 1300~nm. 
The 1300~nm pulse is spectrally filtered to remove any residual pump light or any other colors made by the OPA.
A quarter waveplate~(Thorlabs~AQWP05M-1600) is used to produce circularly polarized laser pulses, which are then focused with a $f=750$~mm CaF$_2$ lens.
As shown in Fig.~\ref{fig::ExpSetup}, a dichroic mirror~(R1300/T400-550) is used to steer the beam into a vacuum chamber.
The streaking laser field is combined with the FEL beam using a silver mirror with a 2~mm drilled hole, and both pulses come to a common focus in the interaction reagion of a co-axial velocity map imaging~(c-VMI) apparatus~\cite{li_co-axial_2018}.
To generate a reference measurement, the streaking laser was intentionally mistimed every 11 XFEL shots.
The c-VMI used here is designed for angular streaking applications, and the device is described in Ref.~\cite{li_co-axial_2018}.
The x-ray focal spot is approximately $\sim55~\mu$m diameter~(FWHM), while the streaking laser focus is substantially larger,$\sim110~\mu$m diameter.
A target gas is introduced via a molecular beam source. 
For the two x-ray photon energies considered in the main text, we use neon as the target for 905~$\rm{eV}$ pulses and CO$_2$ as the target for 570~$\rm{eV}$ pulses. 

Photoelectrons produced by two-color ionization are extracted opposite to the laser propagation direction, as shown in Fig.~\ref{fig::ExpSetup}.
This geometry is used to minimize any signal from scattered x-ray photons. 
Extracted electrons are detected with a microchannel plate detector coupled to a P43 phosphor screen.
The phosphor screen is imaged onto a high-speed CCD camera~(Opal1k) via the 2 mm holey mirror which couples the streaking laser into the chamber, and through the dichroic mirror. 
The CCD camera records images of the phosphor screen at the repetition rate of the accelerator, 120~Hz.

\begin{figure}
    \centering
    \resizebox{0.5\columnwidth}{!}{
    \includegraphics{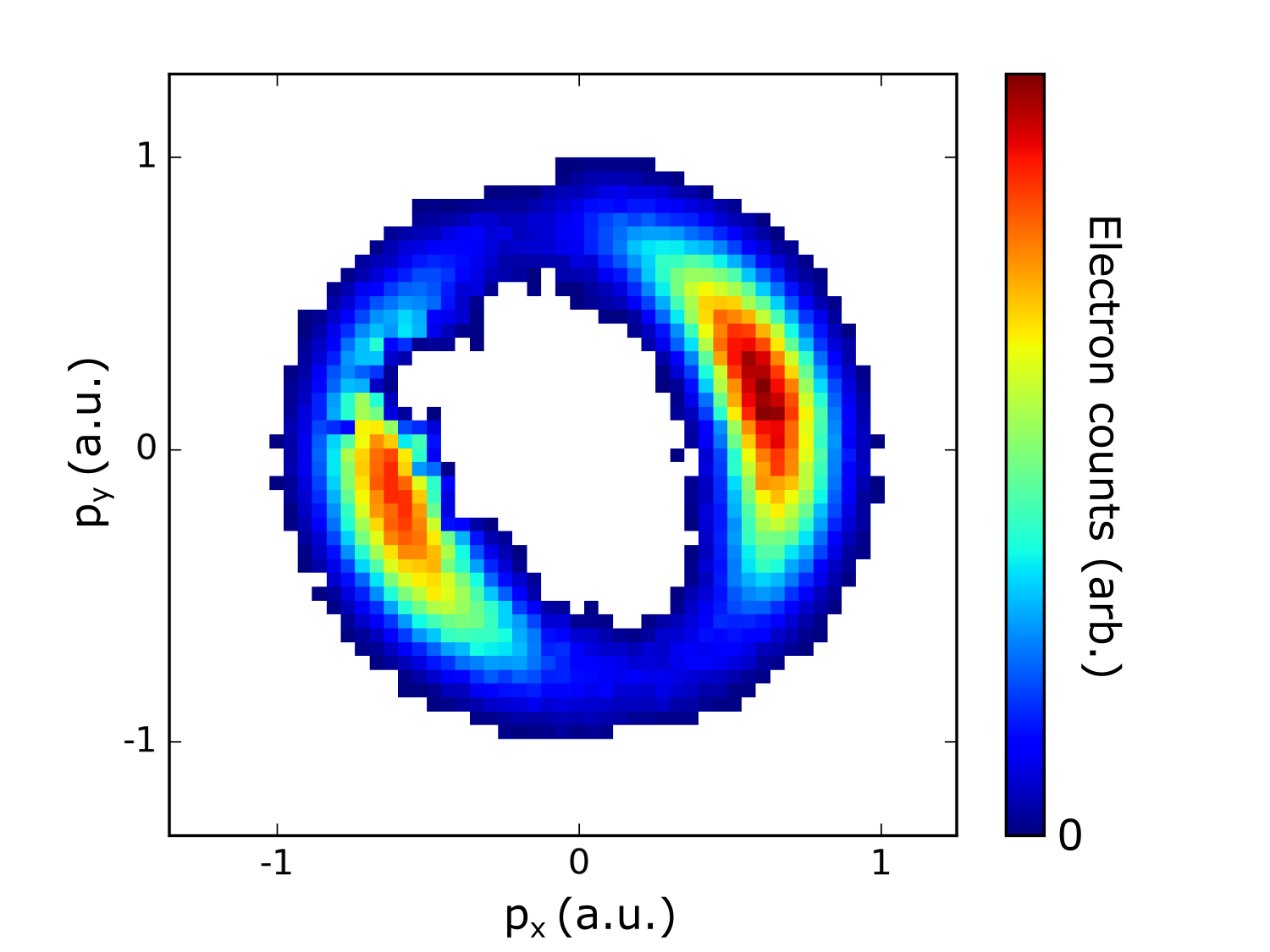}}
    \caption{Two-dimensional projection of the photoelectron momentum distribution generated by the IR laser in CO\textsubscript{2} gas jet. This data was integrated over $4\times10^{4}$ laser shots. The ratio of ionization along the major axis to the minor axis is 10:1, which corresponds to an ellipticity of $\epsilon \sim 0.95$.}
    \label{fig::SFI}
\end{figure}

We can diagnose the streaking laser polarization using the photoelectron momentum distribution from strong-field ionization of a xenon target. 
Figure~\ref{fig::SFI} shows the two-dimensional projection of the photoelectron momentum distribution produced by ionization with the 1300~nm laser.
We can estimate the ellipticity of the streaking laser by comparing the ionization probability along the major and minor axes of the laser polarization.
Using ADK theory the differential ionization rate shown in Figure~\ref{fig::SFI} gives a rough estimate of the ellipticity of $\epsilon \sim 0.95$ for the IR streaking laser.

\subsubsection{Laser X-ray Timing Stability}
\begin{figure}
    \centering
    \resizebox{0.99\columnwidth}{!}{
    \includegraphics{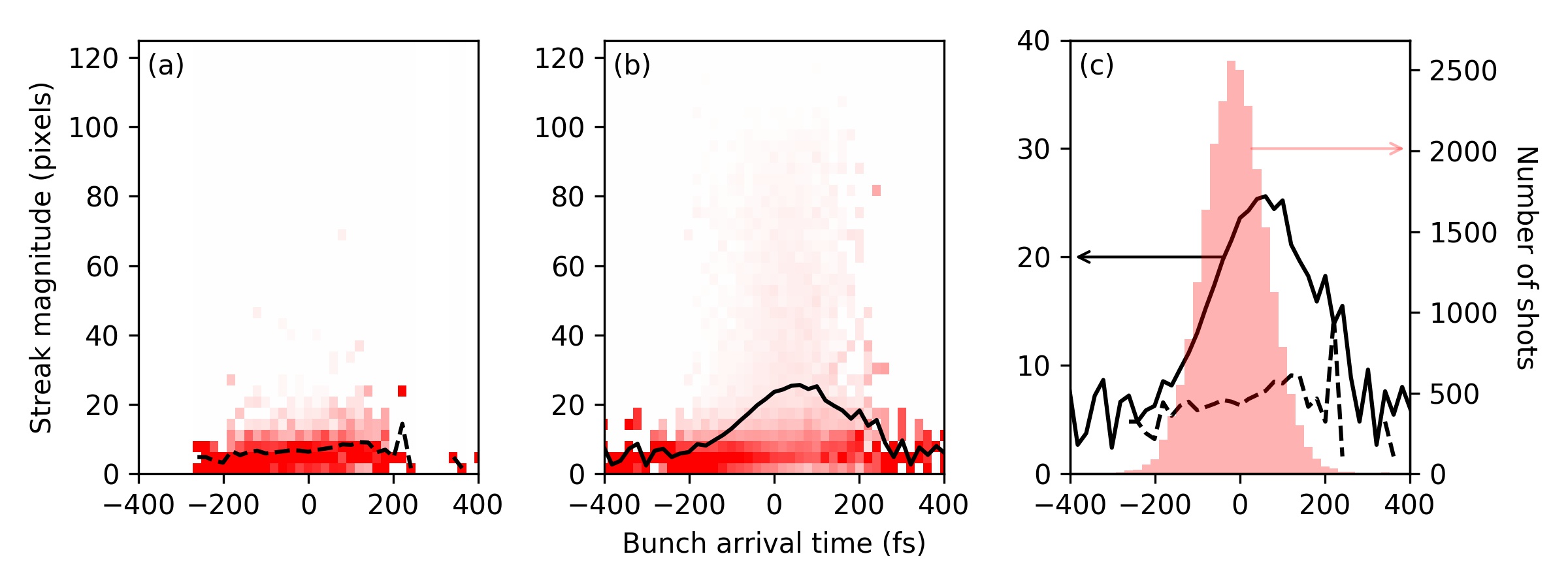}}
    \caption{Distributions of the extracted streak magnitudes as a function of the bunch arrival time with (a) a mistimed streaking laser and (b) a co-timed streaking laser. The black curves show the average streaking magnitude. (c) Comparison of the streaking magnitude's dependence on the bunch arrival time (black, left axis) and the distribution of bunch arrival times (red, right axis).}
    \label{fig::Timing}
\end{figure}
Synchronization between the attosecond X-ray pulses and the streaking pulses is a persistent challenge in our measurement.
Measurements at the LCLS use conventional feedback techniques to stabilize the optical laser pulse arrival time relative to a radio frequency~(RF) reference from the accelerator~\cite{glownia_time-resolved_2010}. 
The primary sources for the temporal jitter include thermal effects, RF noise, and energy jitter in the electron bunch.
The electron bunches accumulate additional temporal jitter as they propagate along the acceleration and bunch compression chain. 
This energy jitter is directly transformed into a timing jitter in the magnetic chicane bunch compressors. 
Using a transverse cavity just before the electron beam dump, we can measure the shot-to-shot variation between the arrival time of the electron bunch and the RF reference.
Figure~\ref{fig::Timing} shows the distribution of measured streaking magnitudes for each shot as a function of the bunch arrival time~(BAT).
The peak of the streaking magnitude distribution is offset from the nominal 0~fs BAT by $\sim$60~fs due to a slight mistiming of the streaking laser.
We observe streaking over a $\sim$300~fs range of BATs.
This range is affected by both the streaking laser pulse width and the streaking laser timing jitter with respect to the RF reference.

\subsection{Photoelectron Data Analysis Procedure}
\begin{figure}
    \centering
    \resizebox{0.99\columnwidth}{!}{
    \includegraphics{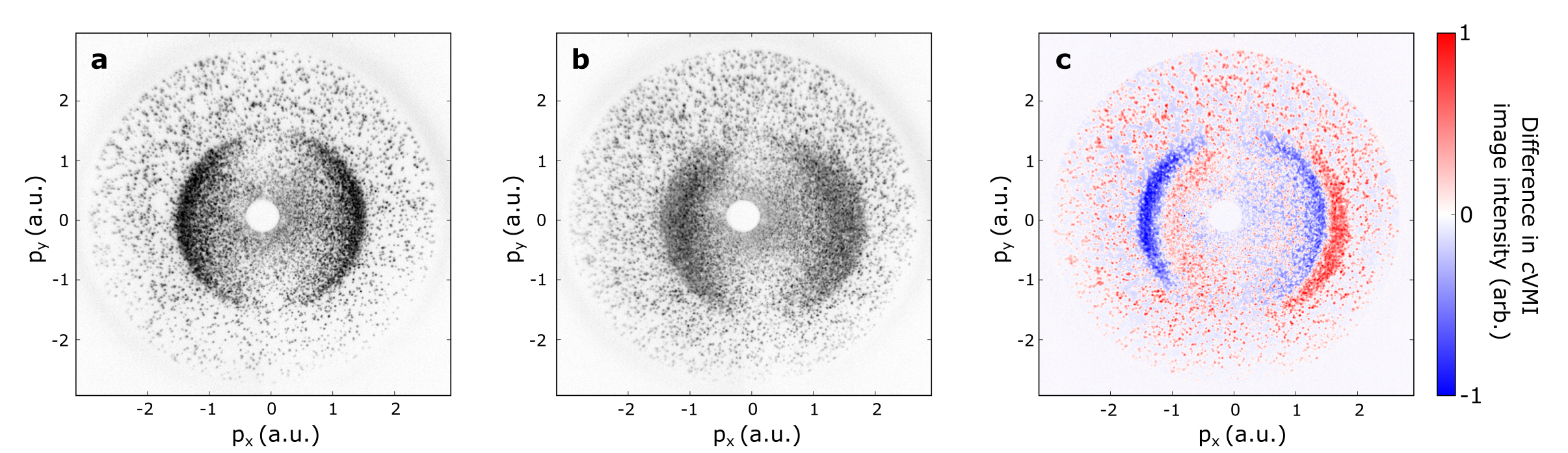}}
    \caption{Measurement of photoelectrons by the c-VMI. (a) raw, unprocessed image of a single-shot measurement of neon 1s photoelectrons ionized by a 905 eV ESASE pulse without the streaking laser. (b) raw single-shot image of neon 1s photoelectrons measured at the same photon energy in the presence of the circularly polarized streaking field, shifting the momentum distribution of the photoelectrons and encoding the properties of the ESASE pulse. This shift is made clear in panel (c), showing the difference in measured electron counts between the shot shown in panel (b) and a background image constructed from multiple measurements of the unstreaked photoelectron distribution.}
    \label{fig::vmi_ims}
\end{figure}
We select shots to analyze based on the single shot pulse energy, streaking angle, and streaking amplitude.
Pulse energy is measured in the standard non-intrusive method employed at the LCLS, by recording the flourescence induced by the x-ray pulse in a small (0.02--1.2 Torr) pressure of nitrogen~\cite{moeller_photon_2011}. 
Filtering on pulse energy ensures the analyzed shots have a high number of electron counts, which provides better quality reconstructions.
A number of complementary techniques provide real-time information on the shot-to-shot amplitude and direction of the momentum shift from the streaking laser. To process the shots, we identify these two streaking coordinates using a genetic optimization algorithm which approximates the centroid of the shifted electron distribution, by maximizing the electron density falling in an open ring shape as the shape is shifted across the c-VMI image~\cite{storn_differential_1997}. 
For the reconstruction, we select shots that are streaked perpendicular to the x-ray polarization, within an angular acceptance of $\sim40^{\circ}$/$\sim65^{\circ}$ for the 905~$\rm{eV}$ and 570~$\rm{eV}$ shots respectively. 
This is a result of the 6~$\rm{mm}$ hole in the center of the detector \cite{li_co-axial_2018} and the angular distribution of the photoelectrons produced by the two-color ionization, which approaches a $\cos^2\theta$ dipole distribution (see Fig.~\ref{fig::vmi_ims}). Shots streaked parallel to the x-ray polarization suffer from a greater distortion of signal because the corresponding momentum shift is along the axis of the highest electron density, maximizing the electron signal lost by virtue of `falling' into the hole. The converse is true for shots streaked perpendicular to the x-ray polarization.
We have verified that the measured pulse duration is not correlated with either pulse energy or streak angle. 

Single-shot CCD images, representative examples of which are shown in Fig. \ref{fig::vmi_ims}, are analyzed in the following way.
First, a quadrant dependent gain correction is applied to the image to account for the variation in camera gain between the four quadrants of the detector, and we shift the center of the image to match the symmetry center found from unstreaked images.
Then we apply a median filter with a neighborhood size of 19~pixels, and the resulting image is convolved with a Gaussian filter~($\sigma=$25~pixels).
The filtered image is then downsized from the initial 1024$\times$1024 image to 32$\times$32 pixels. 
A background image, primarily consisting of low energy electrons at the center of the detector, was obtained from the unstreaked images. This background is subtracted from the downsized image. Incomplete suppression of this electron background for the 905 eV pulse measurements results in an small artifact in the time domain pulse reconstruction (a $<10\%$ feature appearing exactly half the streaking laser period away from the measured pulse). This is removed post-processing. Subtraction of this background from a single shot image inevitably leads to small negative values in some pixels.
Following background subtraction, we threshold the image by the absolute value of the most negative pixel intensity of the single shot image.

\subsection{Reconstruction Algorithm}
The pulse reconstruction algorithm is described in our previous publication~\cite{li_characterizing_2018}.
The algorithm is based on the forward propagation of a basis set. 
We assume that the photoionized electron wavepacket~(EWP) can be described by a set of basis functions, $\alpha_n$:
\begin{equation}
    \psi(t) = \vec{E}_X(t)\cdot\vec{d}(\vec{p}-\vec{A}(t)) = \sum_n c_n~\alpha_n(t).
    \label{Eqn::EWP}
\end{equation}
The EWP,~$\psi(t)$, is given by the product of the x-ray electric field,~$\vec{E}_X(t)$, and the dipole moment,~$\vec{d}(\vec{p})$, describing the x-ray photoionization process, and $\vec{A}(t)$ is the vector potential of the streaking laser field. 
The two-color photoelectron momentum distribution for an arbitrary x-ray pulse can be calculated in the strong field approximation~(SFA)~\cite{kitzler_quantum_2002}:
\begin{eqnarray}
    b(\vec{p}) &=& \int_{-\infty}^{\infty}\vec{E}_{X}(t)\cdot\vec{d}(\vec{p}-\vec{A}(t))~e^{-i\Phi(t)}~dt, \label{Eqn::SFA} \\
    \Phi(t) &=& \int_t^{\infty} dt^{\prime}\left\{ \frac{(\vec{p}-\vec{A}(t^{\prime}))^2}{2}+I_p \right\},
\end{eqnarray}
$b(\vec{p})$ is the probability amplitude for observing an electron with momentum $\vec{p}$, and $I_p$ is the binding energy of the ionized electron.
We assume the dipole moment is described by a hydrogenic model: $\vec{d}(\vec{p})=\vec{p}/(|\vec{p}|^2 - 2I_p)^3$.
Each basis function, $\alpha_n(t)$, will produce a probability amplitude, $a_n(\vec{p})$ for observing an electron with momentum $\vec{p}$ according to:
\begin{eqnarray}
    a_n(\vec{p}) &=& \int_{-\infty}^{\infty}\vec{\alpha}_{n}(t)\cdot\vec{d}(\vec{p}-\vec{A}(t))~e^{-i\Phi(t)}~dt.
\end{eqnarray}
The total probability amplitude, $b(\vec{p})$, for the EWP in Eq.~\ref{Eqn::EWP} is given by,
\begin{equation}
    b(\vec{p})=\sum_n c_n a_n(\vec{p}).
\end{equation}
The measured photoelectron momentum distribution is given by the 
\begin{equation}
    B(\vec{P})=\int dp_z\left|b(\vec{p})\right|^2 = \sum_{nm} c^*_m c_n \int~dp_z a^*_m(\vec{p})a_n(\vec{p}),
\label{supp_eq:Bp}
\end{equation}
where $\vec{P}$ is the two-dimensional vector describing the projected momentum.
While Eqs.~\ref{Eqn::SFA}--\ref{supp_eq:cost_function} are general for any choice of basis function, in this work we choose to construct the basis functions $\alpha_n$ by forward-propagating a set of x-ray pulses described by the von Neumann functions, which are a joint time-frequency basis:
\begin{equation}
    \alpha_{ij}(t) =\left(\frac{1}{2\alpha\pi}\right)^{1/4}\exp\bigg[-\frac{1}{4\alpha}(t-t_j)^2-it\omega_i\bigg] .
\end{equation}
where $\alpha$ is a constant specified by the von Neumann lattice range and size, and the lattice points~($t_i,w_j$) are distributed on a 6$\times$6 grid~\cite{fechner_von_2007}.
The total time window for the von Neumann lattice is chosen to be one period of the streaking laser field~(4.3~fs for a central frequency of 1.3~$\mu$m).
The number of lattice points is selected empirically to produce a reliable reconstruction, which captures the intricate structures of sub-femtosecond pulses and at the same time does not suffer from high frequency artifacts introduced by fitting too many basis functions for the small image size.
A 6$\times$6 von Neumann lattice corresponds to 36 points in the time domain representation, giving a temporal spacing of 120~$\rm{as}$ for the temporal reconstruction. 

\begin{figure}
    \centering
    \resizebox{0.99\columnwidth}{!}{
    \includegraphics{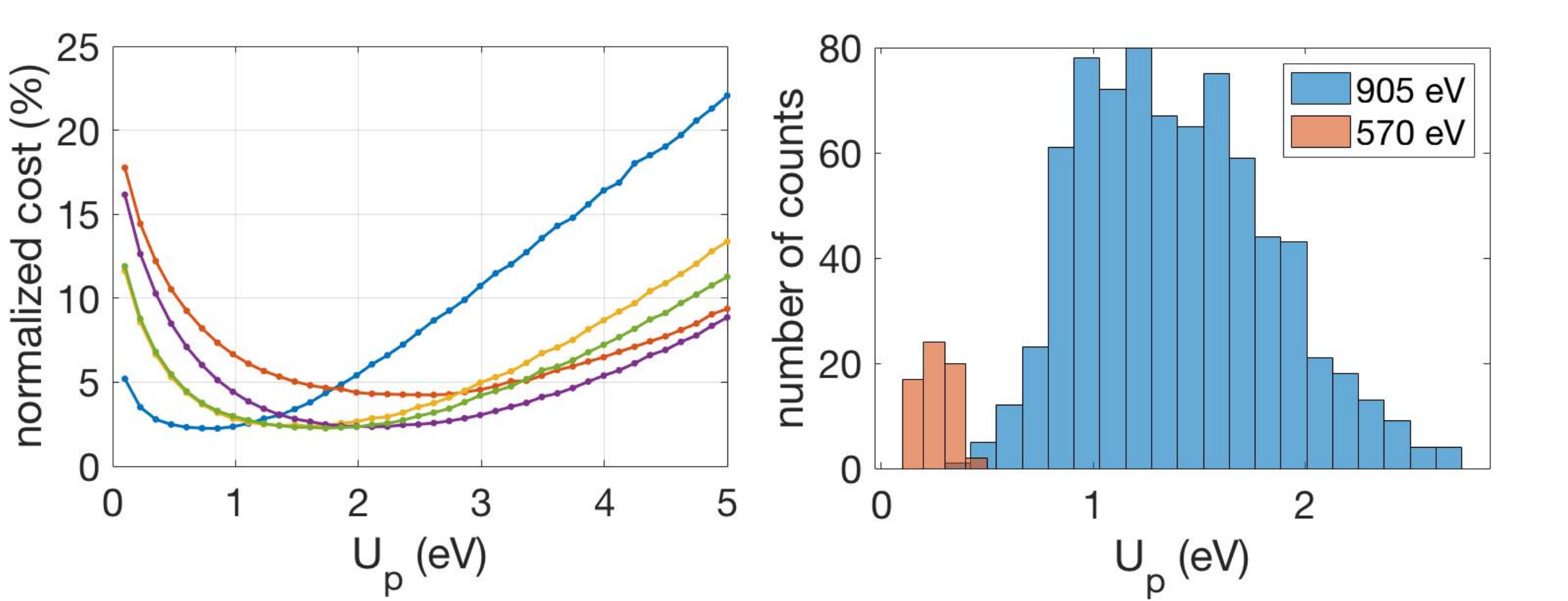}}
    \caption{Left: Examples of $U_p$ scan for five different shots. Normalized cost is defined as $L_G$ in Eq.~\ref{supp_eq:cost_function} divided by the sum of image intensity squared. Right: distribution of $U_p$ for the analyzed shots.}
    \label{fig::Upscan}
\end{figure}

We use a nonlinear minimization routine to fit the measured single-shot electron distribution to the basis set constructed from each von Neumann function.
The likelihood estimator $L$ measures the agreement between the simulated data from Eq.~\ref{supp_eq:Bp} and the actual data, $D$:
\begin{eqnarray}
    L_G &=& \sum_i \left|D_i - B(\vec{P}_i)\right|^2, \\
    L_P &=& \sum_i \left\{B(\vec{P}_i) - D_i\mbox{ln}[B(\vec{P}_i)] + \mbox{ln}[D_i!]\right\},
\label{supp_eq:cost_function}
\end{eqnarray}
where $i$ indexes the pixels of the measurement. 
$L_G$ is the maximum likelihood estimator in the case of Gaussian distributed noise, whereas $L_P$ is the maximum likelihood estimator in the case of Poisson distributed noise.
Both likelihood estimators were tested and provide similar results. 
Use of the von Neumann representation affords a sparse representation of the x-ray pulse, which assists in the robustness of the nonlinear fitting routine. 

As described above, the relative phase between the x-ray pulse and the streaking pulse is random, and we do not have knowledge of the instantaneous value of the vector potential amplitude, or the ponderomotive potential, $U_p$, responsible for the momentum shift. 
As shown in~\cite{li_characterizing_2018}, we find the optimal $U_p$ by scanning the likelihood estimator, Eq.~\ref{supp_eq:cost_function}, as a function of $U_p$ with a range from 0.5 to 5~$\rm{eV}$. 
The likelihood estimator varies smoothly with $U_p$, showing a clear minimum which identifies the optimal $U_p$ (see Fig.~\ref{fig::Upscan}).

The basis images (Eq.~\ref{supp_eq:Bp}) are calculated on a $64\times64$ grid and downsized to 32$\times$32 to prevent numerical artifacts, and then convolved with a Gaussian filter with $\sigma$ equivalent to the convolution strength of 25~pixels on a 1024$\times$1024 grid, to match the filtering performed on the data.

\begin{figure}
    \centering
    \resizebox{0.6\columnwidth}{!}{
    \includegraphics{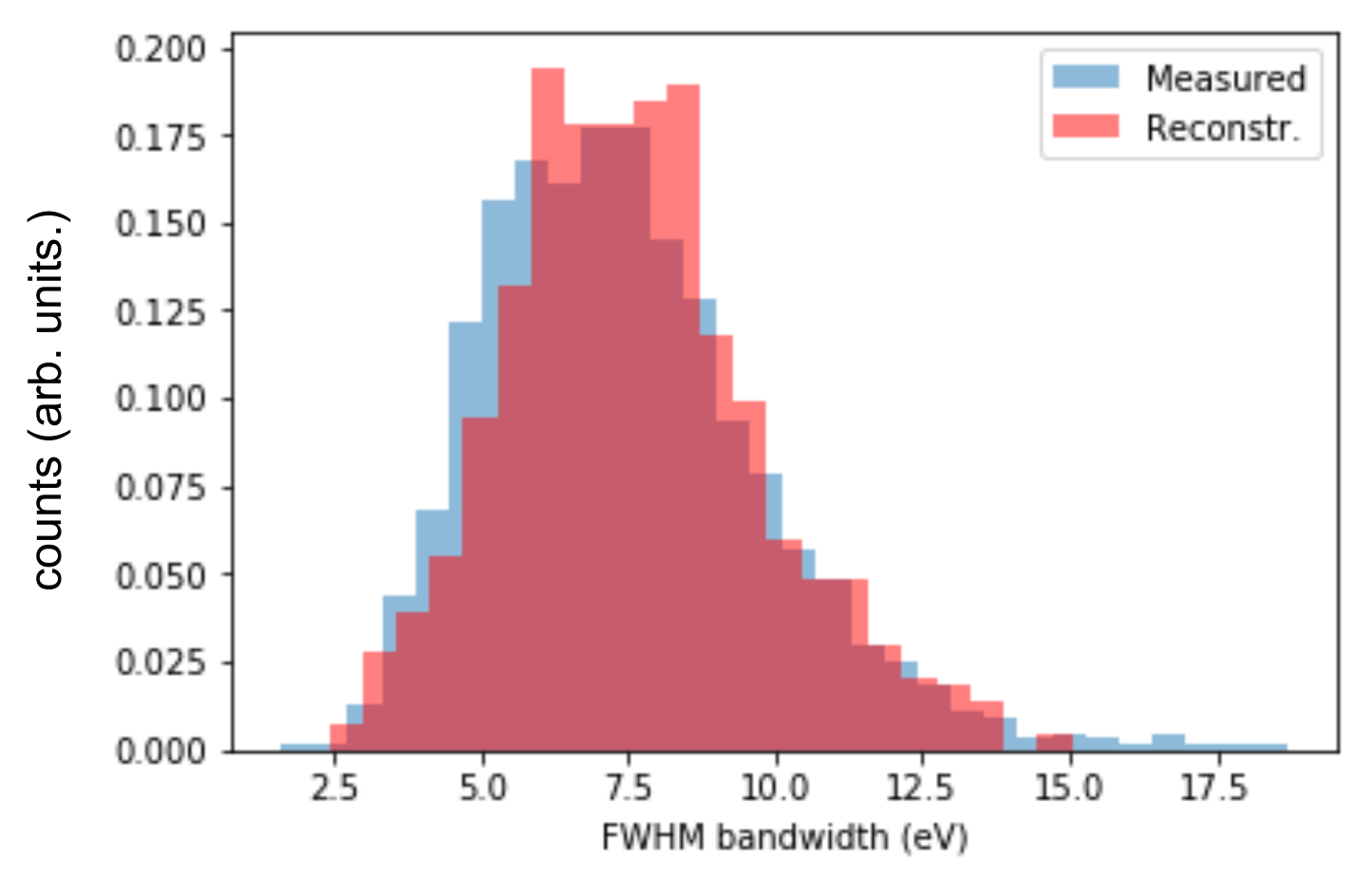}}
    \caption{Bandwidth distributions from spectrometer measurements at 905~$\rm{eV}$ (blue) and reconstructed shots from streaking measurements at 905~$\rm{eV}$.}
    \label{fig::bwcomparison}
\end{figure}

The nonlinear fitting search is initiated by random numbers for the coefficients based on Monte Carlo sampling. 
For each shot, we run 20 iterations of the reconstruction algorithm with different initial values and check the convergence by comparing each solution. 
The time domain profile of the x-ray field is given by,
\begin{equation}
    E_X(t) = \sum_n c_n \alpha_n(t),
\end{equation}
where $c_n$ are the optimal coefficients returned by the algorithm. 
We then take the Fourier transform of the time-domain x-ray field profile to obtain the spectral domain profile.

\begin{figure}
    \centering
    \resizebox{0.55\columnwidth}{!}{
    \includegraphics{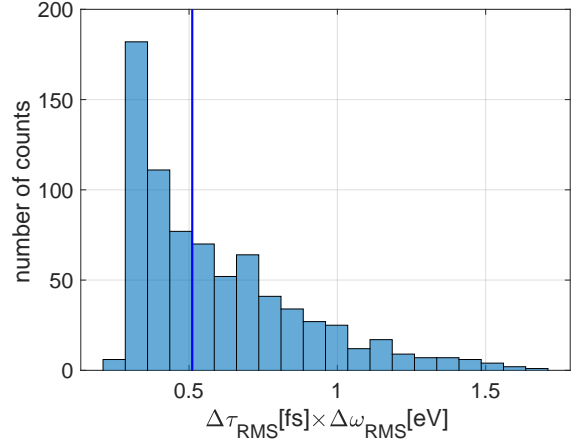}}
    \caption{Distribution of time-frequency RMS bandwidth product of 905~$\rm{eV}$ data. The vertical blue line indicates the median of the distribution at 0.51.}
    \label{fig::bwproduct}
\end{figure}

Note that the spectral domain is sensitive to spiky structures in the time domain profile which are artifacts of the algorithm due to Poisson noise. 
To mitigate this problem, we apply a Gaussian filter to the time domain field profile with a $\sigma_t=170~\rm{as}$ for 570~$\rm{eV}$ and $\sigma_t=120~\rm{as}$ for 905~$\rm{eV}$. 
The Gaussian filter width is comparable to the basis function lattice grid size (120~$\rm{as}$). 
With this Gaussian filter the statistical distribution of the FWHM bandwidth obtained from the reconstruction closely matches the spectrometer data (see Fig. .~\ref{fig::bwcomparison}), which provides an independent benchmark for the reconstruction method.

Figure~\ref{fig::bwproduct} shows the distribution of time-frequency bandwidth product measured from the reconstructed pulses. For a Fourier-transform limited Gaussian pulse the time-frequency RMS bandwidth product should be 0.32 and the average time-bandwidth product is within a factor two of this value.

\subsubsection{Quantification of the single-shot experimental uncertainty}
\begin{figure}
    \centering
    \resizebox{0.99\columnwidth}{!}{
    \includegraphics{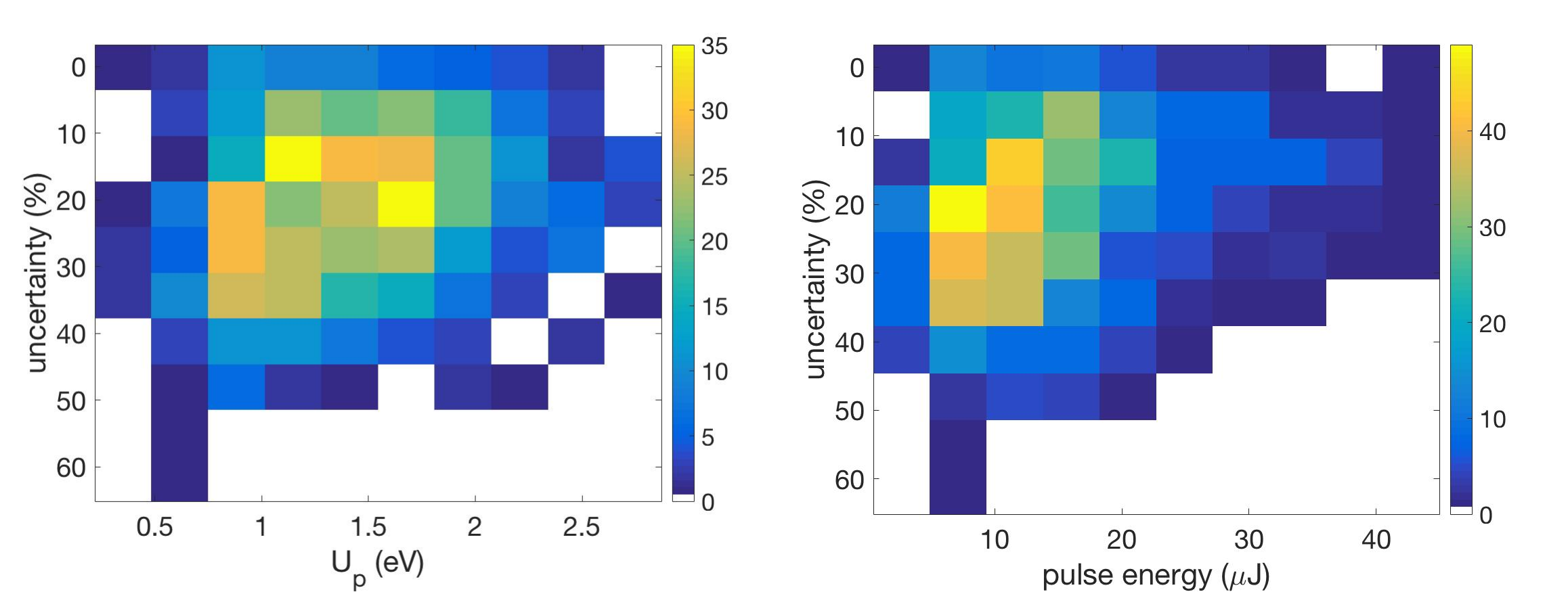}}
    \caption{Distribution of pulse duration uncertainty as a function of pulse energy and streaking amplitude.}
    \label{fig::uncertainty_correlation}
\end{figure}
The reconstruction algorithm can converge to different solutions for a given pulse due to the finite number of counts in each shot and the finite resolution of the VMI. These different solutions all match the measured data very accurately, with an integrated error smaller than 3\%. 
Therefore we can employ the statistical distribution of these independent reconstructions to quantify the most probable solution as well as the experimental uncertainty on the pulse profile and pulse duration.

 To obtain the most probable solution, we calculate the inner product between each of the 20 solutions (i.e. $<I_i,I_j>$ with $i,j=$~1,2,...,20) for each shot. We then identify the most probable solution as the one with the highest inner product, shown as the red curve in main text Fig.~2. The standard deviation of the inner product of these solutions is a measure of the uncertainty on the overall pulse profile and it has an average value of 6\%.

To obtain an estimate of pulse duration uncertainty we calculate the standard deviation of the pulse duration found by the 20 solutions. 
We define the uncertainty to be ${\sigma\left(\Delta\tau_{\rm{RMS}}\right)}/{\Delta\tau_{\rm{RMS}}}$, where $\sigma$ is the standard deviation. 
The median of the uncertainty distribution is 20\% (28\%) for 905~$\rm{eV}$ (570~$\rm{eV}$).
Figure~\ref{fig::uncertainty_correlation} shows the pulse duration uncertainty correlates with the streaking amplitude and the pulse energy, indicating that the uncertainty is smaller for higher streaking amplitude as well as for higher pulse energy. The dependence on the streaking amplitude comes from the decreasing angular resolution corresponding to small radius on the detector. In the limiting case where the streaking amplitude is 0, there is complete degeneracy across the time axis. The dependence on higher pulse energy can be explained by stronger signal.

\section{Supplemental Text}
\subsection{Two-Color Pulse Properties}
\begin{figure}
    \centering
    \resizebox{0.7\columnwidth}{!}{\includegraphics{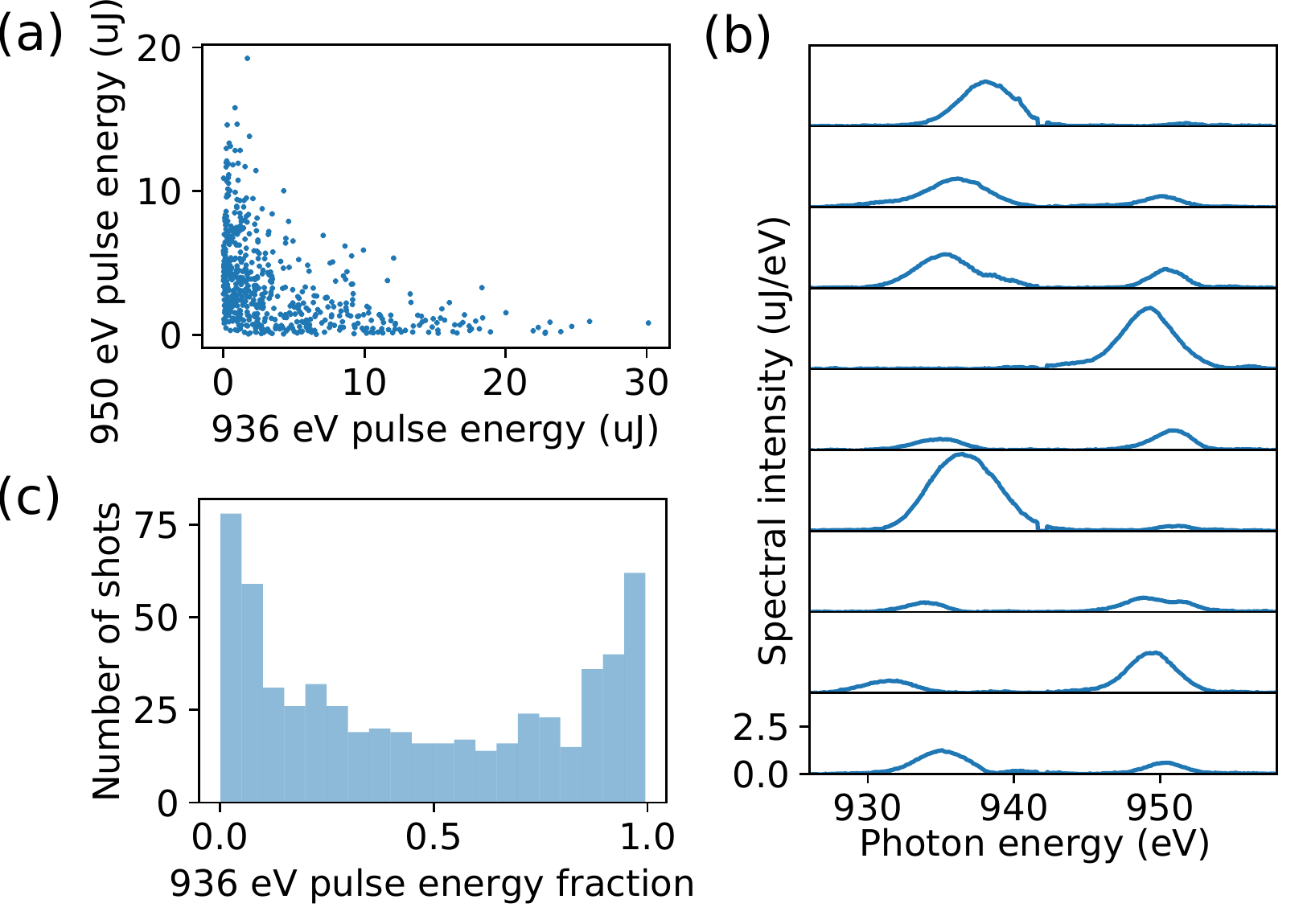}}
    \caption{Single-shot spectral measurements for the two-color experiment. (a) Single-shot measurement of the intensity of each pulse. (b) Representative single-shot spectra for the two-color setup. (c) Histogram of the fraction of the total pulse energy in each pulse.}
    \label{fig::TwoColorSpectra}
\end{figure}
The characteristic intensity jitter of single-spike sub-fs FELs is also observed in the double-pulse operation. Figure~\ref{fig::TwoColorSpectra}~(a) shows a scatter plot of the pulse energy contained in the two pulses from shot-to-shot.
The pulse energy in each pulse has close to 100\% fluctuations, and the intensities of the two pulses are anti-correlated, a well known property of the split undulator method~\cite{Lutman2C}. This is because when running the system as close as possible to saturation, a large increase in the energy of the first pulse spoils the energy-spread of the electron beam, causing the second pulse to be weaker. One could improve the relative pulse stability at the cost of a reduced peak power by interrupting the gain of the first pulse earlier and reducing the effect of the first pulse on the second.

Figure~\ref{fig::TwoColorSpectra}~(b) shows a few examples of single shot two-color spectra, while Fig.~\ref{fig::TwoColorSpectra}~(c) is a histogram of the relative intensities of the two pulses. While the source has significant intensity jitter, the pulse energy is so large that the properties can be measured on a single-shot basis for all the shots with non-destructive diagnostics. Therefore one could apply this method to time-resolved pump/probe experiments and sort the data based on single-shot measurements of the pulse properties.

\subsection{Machine Parameter Correlations}
\begin{figure}
    \centering
    \resizebox{0.99\columnwidth}{!}{
    \includegraphics{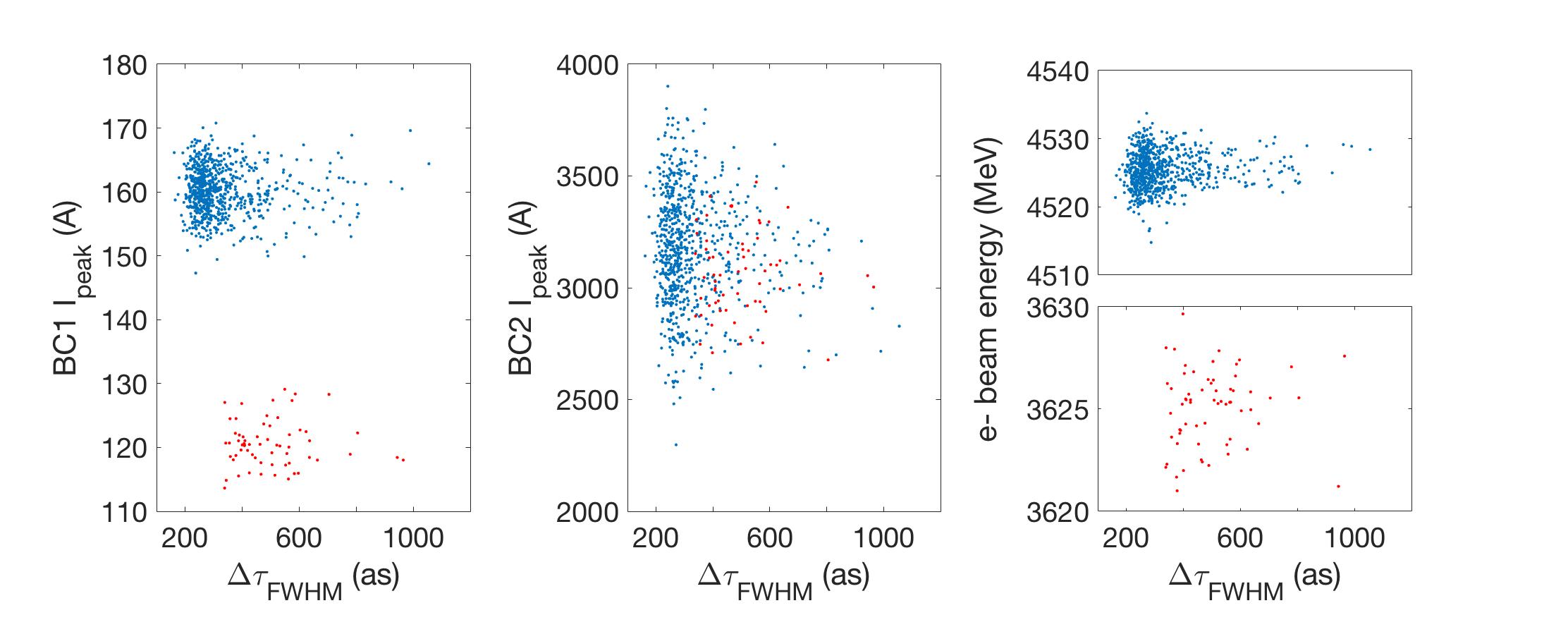}}
    \caption{Scatter plots of machine parameters and the retrieved pulse durations. The blue dots are for the neon data at 905~$\rm{eV}$, and red for the CO$_2$ data at 570~$\rm{eV}$. }
    \label{fig::ebeam_corr}
\end{figure}
We have examined the correlation between the retrieved pulse durations and a number of critical machine parameters: the peak electron beam current measured at bunch compressors 1 and 2 (BC1 and BC2) and the electron beam energy after acceleration. 
Figure ~\ref{fig::ebeam_corr} shows that there is no clear correlation between either of these parameters and the measured pulse duration (the calculated correlation coefficients are all below 0.2).



\bibliographystyle{science.bst}
\bibliography{referencesJPC.bib,biblio.bib}